\documentclass[aps,pra,preprint,amsmath,amssymb]{revtex4-1}

\usepackage{amsmath}
\usepackage{amssymb}
\usepackage{graphicx}
\usepackage[normalem]{ulem}
\usepackage{bbm}
\usepackage{bm}
\usepackage[caption=false]{subfig}
\usepackage{color}
\usepackage{dsfont}

\DeclareMathOperator{\tr}{tr}

\newcommand{\phid}{\phi^\dag}

\newcommand{\ctheta}{\vartheta}
\newcommand{\PHI}{\mathbf{\Phi}}
\newcommand{\MR}{\mathbf{R}}
\newcommand{\MG}{\mathbf{G}}
\newcommand{\MGamma}{\bm{\Gamma}}
\newcommand{\Q}{\mathbf{q}}
\newcommand{\X}{\mathbf{x}}
\newcommand{\vP}{\mathbf{p}}
\newcommand{\vGamma}{\dot{\Gamma}}
\newcommand{\vPHI}{\dot{\PHI}}

\newcommand{\vrho}{\dot{\rho}}
\newcommand{\vu}[1]{\dot{u}_{#1}}
\newcommand{\vn}[1]{\dot{n}_{#1}}

\newcommand{\vZctheta}{\dot{Z}_\ctheta}
\newcommand{\vYm}{\dot{Y}_m}
\newcommand{\vZphi}{\dot{Z}_\phi}

\newcommand{\bk}{b_k}

\newcommand{\ddX}{\mathrm{d}^d\X}

\newcommand{\ddQ}{\mathrm{d}^d\Q}
\newcommand{\dtau}{\mathrm{d} \tau}

\begin{document}

\title{Thermodynamics of Bose gases from functional renormalization with a hydrodynamic low-energy effective action}

\author{Felipe Isaule}
\author{Michael C. Birse}
\author{Niels R. Walet}
\affiliation{Theoretical Physics Division, School of Physics and Astronomy, The University of Manchester, Manchester, M13 9PL, United Kingdom}

\begin{abstract}
The functional renormalization group for the effective action is used to construct an effective hydrodynamic description of weakly interacting Bose gases. We employ a scale-dependent parametrization of the boson fields developed previously to start the renormalization evolution in a Cartesian representation at high momenta and interpolate to an amplitude-phase one in the low-momentum regime. This technique is applied to Bose gases in one, two and three dimensions, where we study thermodynamic quantities such as the pressure and energy per particle. 
The interpolation leads to a very natural description of the Goldstone modes in the physical limit, and compares well to analytic and  Monte-Carlo simulations at zero temperature.
The results show that our method improves aspects of the description of low-dimensional systems, 
with  stable results for the superfluid phase in two dimensions and even in one dimension.
\end{abstract}

\maketitle

\section{Introduction}
\label{sec:Intro}

There is a long standing interest in finding a consistent theoretical description of Bose gases, see Ref.~\cite{pitaevskii_bose-einstein_2016} for a complete history.
At the simplest level, mean-field theory gives a  qualitative description of three-dimensional systems in the superfluid phase \cite{bogoliubov_n_theory_1947},
but to obtain accurate results the full effects of fluctuations must be included. This is particularly important  in the infrared (IR) where
bosons condense~\cite{andersen_theory_2004}.
These fluctuations can in principle be treated systematically using a perturbative expansion, but
this is plagued by order-by-order IR divergent contributions \cite{gavoret_structure_1964}.
There are strong cancellations between these divergences resulting in finite thermodynamic 
properties \cite{nepomnyashchii_contribution_1975,pistolesi_renormalization-group_2004}. 
These cancellations can be lost  if expansions are truncated at finite order.

The impact of fluctuations is greater in low-dimensional systems \cite{al_khawaja_low_2002}.
Their effects can be so strong that they destroy the long-range-order (LRO) in dimensions below three, thus suppressing Bose-Einstein condensation \cite{mermin_absence_1966}. In a homogeneous two-dimensional gas, condensation is only possible at zero temperature. Nonetheless, superfluid behavior can
still be present at finite temperatures where correlation functions show power-law decays or 
quasi-long-range-order (QLRO) \cite{posazhennikova__2006}.
Fluctuations are even more important in one dimension, where a homogeneous gas does not condense
at any temperature. Even there, superfluidity is still possible at zero temperature \cite{cazalilla_one_2011}.
As shown in Refs.~\cite{buchler_superfluidity_2001,astrakharchik_motion_2004}, the one-dimensional gas
shows superfluid features in the weakly-interacting limit at zero temperature. However, unlike
its two- and three-dimensional counterparts, the one-dimensional system does not show a phase transition, 
and is in the normal phase at any non-zero temperature.
Moreover, the superfluid fraction at zero temperature continuously decreases as the gas becomes more strongly interacting, 
until it disappears in the Tonks-Girardeau limit of impenetrable bosons  \cite{girardeau_relationship_1960}.

One natural way to describe Bose gases is with a field theory of non-relativistic interacting complex boson fields~\cite{stoof_ultracold_2009}, 
which can then be tackled with field-theoretical methods, for instance using diagrammatic expansions. All such techniques rely on calculating loop diagrams, 
where it is difficult to treat infrared (IR) fluctuations using 
the Cartesian representation where the boson fields
are decomposed into  real (longitudinal) and imaginary (Goldstone) parts.
The strong coupling between longitudinal and Goldstone fluctuations
produces IR divergent terms even at the level of the one-loop corrections. 
Although these divergences cancel to leave a finite result, they require a sophisticated treatment~\cite{pistolesi_renormalization-group_2004}. 
For instance, Nepomnyashchii and Nepomnyashchii were
able to compute the correct behavior of the anomalous self-energy in the IR by employing a self-consistent analysis of the perturbative expansion in order to
cancel the divergent diagrams~\cite{nepomnyashchii_contribution_1975}. 
An alternative approach, employing dimensional regularization, can be found in  Ref.~\cite{andersen_theory_2004}.
However, a more convenient method to avoid these problems is to employ
an amplitude-phase (AP) representation for the boson fields in the IR, as
in Popov's hydrodynamic effective theory \cite{popov_functional_1987}, where the fields are decomposed in radial and phase (Goldstone)
parts. In this representation the Goldstone fields appear only in interactions coupled through their derivatives, so 
there are no IR divergences  and perturbation theory can be used without requiring delicate cancellations
(for a complete discussion on how both representations are connected we refer to Refs.~\cite{pistolesi_renormalization-group_2004,dupuis_infrared_2011}).
The AP representation is now widely used in modern calculations to describe the low-energy regime
of Bose gases, particularly in low dimensions \cite{al_khawaja_low_2002,prokofev_two-dimensional_2002,prokofev_weakly_2004,mora_extension_2003}.
With these developments, weakly-interacting Bose gases are now generally well understood and described (see for example Ref.~\cite{capogrosso-sansone_beliaev_2010}),
increasing the theoretical interest in more complicated related systems such as strongly correlated 
Bose \cite{chevy_strongly_2016} and Fermi gases~\cite{randeria_crossover_2014},
multi-component gases~\cite{pitaevskii_bose-einstein_2016,catani_degenerate_2008,ferrier-barbut_mixture_2014}, among others.

A rather different approach to  Bose gases is the functional renormalization group (FRG) \cite{wetterich_exact_1993,berges_non-perturbative_2002}.
This is a non-perturbative technique where a parametrization of the full effective action of the system is calculated by gradually integrating out the fluctuations of the fields
as a cut-off on the low-momentum modes is lowered. It typically takes the form of a set of flow equations for the couplings in the scale-dependent effective action.
The FRG has mainly been implemented using the Cartesian representation for the fields and it has 
been  successful in describing bulk thermodynamic properties and critical exponents of three-dimensional Bose gases 
\cite{andersen_application_1999,blaizot_non-perturbative_2005,floerchinger_functional_2008,floerchinger_nonperturbative_2009,eichler_condensate_2009,rancon_thermodynamics_2012}.
As a non-perturbative approach, the FRG does not show the IR divergences of perturbation theory \cite{wetterich_functional_2008,dupuis_infrared_2009,sinner_functional_2010},
however it has been argued that the gradient expansion might not be valid in the extreme IR~\cite{dupuis_infrared_2011}. 
In three dimensions all relevant quantities saturate before such small scales are reached, and eventual numerical complications can be avoided by rescaling the scale-dependent couplings.
In contrast, in low dimensions the FRG has been less successful.
In two dimensions, although bulk thermodynamic properties can be  obtained as they quickly saturate, 
the superfluid stiffness shows an unrealistic decay in the IR regime, probably due to the truncation of action, resulting in 
a non-superfluid system at \emph{any} finite temperature~\cite{grater_kosterlitz-thouless_1995, gersdorff_nonperturbative_2001}.
As proposed by Jakubczyk \emph{et al.} \cite{jakubczyk_longitudinal_2017}, this issue may be avoided by fine-tuning the regulator. 
This is  contrary to the spirit of the FRG method, where reliable results ought to be independent of the
choice of regulator. It
suggests that current FRG calculations are not sufficiently robust.
Similar reservations apply to the
critical exponents, which can be extracted only indirectly from a line of pseudo-fixed points~\cite{floerchinger_superfluid_2009,rancon_universal_2012}. 
As expected, such difficulties with obtaining meaningful results from FRG calculations become even more pronounced in one dimension, where the
gradient expansion becomes invalid when the anomalous dimension becomes large~\cite{dupuis_non-perturbative_2007}.

As shown first by Defenu \emph{et al.}~\cite{defenu_nonperturbative_2017}, 
these IR issues can be easily overcome by using the AP representation
for the fields. Working at lowest order of the gradient expansion,
the FRG then recovers a stable superfluid phase in two dimensions.
However, in Ref.~\cite{defenu_nonperturbative_2017}, the authors simply subtract the contribution of Gaussian ultraviolet (UV) fluctuations, which makes such an  approach 
difficult to apply to dynamical systems where the interaction needs to be renormalized.
Such problems are caused by the fact that just as the Cartesian representation is not the best choice in the IR, 
the AP is in general a poor choice in the UV regime. In a previous paper \cite{isaule_application_2018}, we implemented a scale-dependent parametrization of the boson fields that interpolates between
the Cartesian representation in the UV and the AP representation in the IR.
This ``interpolating representation" enables us to treat correctly both Gaussian and Goldstone fluctuations in the UV and IR, respectively. 
In order to test the approach, we studied its application to classical $O(2)$-models in two and three dimensions.
As suggested by different works (see for example Ref.~\cite{capogrosso-sansone_beliaev_2010}), the transition between representations should
be made around the healing scale \cite{pitaevskii_bose-einstein_2016}, so that the AP representation is used in the IR regime dominated by Goldstone fluctuations, whereas the Cartesian representation is used for the fluctuations in the UV, where they can be treated as Gaussian.
We found that if we make the switch-over fast enough, the results are stable, with a sensible behavior of the parameters in the physical limit.
Furthermore,  the flowing couplings correctly switch from  Cartesian behavior in the UV to the expected forms for the hydrodynamic effective action
in the IR.

In the present work we extend this approach to the study of weakly-interacting Bose gases in two and three dimensions both at zero and finite
temperatures, and in one dimension at zero temperature only. We aim to give a consistent description of the thermodynamic properties of these systems
by computing the pressure, entropy and density of the system.
Even though weakly-interacting Bose gases can be described quite well using other approaches~\cite{pitaevskii_bose-einstein_2016,capogrosso-sansone_beliaev_2010}, the main goal of the present article is to develop and analyze techniques that allow us to generate an improved description using the FRG. This will also be relevant to applications of the FRG to related systems, such as paired fermions, relativistic bosonic field theories, etc.

Our paper is organized as follows. In Sec.~\ref{sec:Action} we present our ansatz for the effective action.
In Sec.~\ref{sec:Broken} we give specific details how the flow equations are solved and we summarize our interpolating approach.
In Sec.~\ref{sec:Thermo} we explain how the bulk thermodynamic properties are obtained. 
In. Sec.~\ref{sec:IC} we give the initial conditions, including the renormalization of the interaction.
Finally, in Sec.~\ref{sec:Results} we present our results for one, two and three dimensions.

\section{Effective action}
\label{sec:Action}

We consider a system of bosons weakly interacting through a short-range repulsive potential.
Expressed in terms of the complex boson field $\phi$ and using the imaginary time $\tau=i\,t$, 
the bare action takes the form
\begin{equation}
\mathcal{S}[\PHI]=\int_x \left[\phid\left(-\partial_\tau+\frac{\nabla^2}{2m}+\mu\right)\phi-\frac{g}{2}(\phid\phi)^2\right].
\label{eq:S}
\end{equation}
Here $\int_x=\int_0^\beta\dtau\int\ddX$,  $\beta=1/T$ is the inverse temperature and $\mu$ is the chemical potential.
The repulsive potential has been approximated by a contact interaction with a strength $g$ that is related to the s-wave scattering length (see Sec.~\ref{sec:IC}). 
Here and in the following we express all physical quantities in units where $\hbar=k_B=1$. We have 
also introduced the field vector $\PHI=(\phi,\phid)$.

As in our previous work \cite{isaule_application_2018}, we use the FRG to obtain a flow equation for the scale-dependent effective action $\Gamma_k$ of the system (the generator of one-particle irreducible Green's functions). The dependence on the momentum scale $k$ is introduced by adding a regulator $\MR$ that
suppresses all quantum and thermal fluctuations for momenta $q<k$.
We start the flow at a UV scale $\Lambda$ from the bare action $\Gamma_\Lambda=\mathcal{S}$, and at the end of the flow, for $k=0$, $\Gamma_0$ is the full effective action.
This allows us to extract the thermodynamic properties of the system from the  grand canonical potential.

In this work we parametrize the boson fields $\PHI$ so they change their representation with $k$.
In this case,  the evolution of the action $\Gamma$ as a function of $k$
 is governed by the flow equation \cite{pawlowski_aspects_2007},
\begin{align}
\partial_{k}\Gamma+\vPHI\cdot\frac{\delta \Gamma}{\delta \PHI}=\frac{1}{2}\tr\left[\partial_k \MR(\MGamma^{(2)}-\MR)^{-1}\right] +\tr\left[\vPHI^{(1)} \MR(\MGamma_k^{(2)}-\MR)^{-1}\right],
\label{eq:PawlowskEq}
\end{align}
where  $\partial_{k}\Gamma$ represents the $k$ derivative for constant fields, $\vPHI=\partial_k \PHI$ is the $k$ derivative of the fields, 
$\MGamma^{(2)}$ is the matrix of second functional derivatives with respect to the fields $\PHI$,
and $\vPHI^{(1)}$ is the matrix of the first functional derivatives of $\vPHI$ with respect to the fields. 
The regulator $\MR$ is a diagonal matrix with elements $R_k(q)$. In this work we adopt a frequency-independent exponential regulator in the form
\cite{gersdorff_nonperturbative_2001,canet_optimization_2003}
\begin{equation}
R^{\text{exp}}(\Q)=\frac{Z_m \Q^2/2m}{\exp(\Q^2/k^2)-1},
\label{eq:Rexp}
\end{equation} 
where $Z_m$ is defined below. This commonly used regulator has the benefit of a smooth decay around $q=k$ \cite{gersdorff_nonperturbative_2001,canet_optimization_2003}.

In order to solve flow equation (\ref{eq:PawlowskEq}) we approximate $\Gamma$ using a gradient expansion.
We use an ansatz up to fourth order in the fields and second order in derivatives
\begin{equation}
 \Gamma[\PHI]=\int_x \left[\phid\left(-Z_\phi\partial_\tau+\frac{Z_m}{2m}\nabla^2\right)\phi+\frac{Y_m}{8m}\rho\nabla^2\rho-U(\rho,\mu) \right],
 \label{eq:Gammak}
\end{equation}
where $Z_\phi$, $Z_m$ and $Y_m$ are $k$-dependent renormalization factors and, at the level of truncation used in this work, field-independent.
Since we employ a periodic imaginary time variable $\tau=it$ to describe systems at finite temperature the energy integrals are replaced by sums
over Matsubara frequencies~\cite{stoof_ultracold_2009}. We stress that the term $\frac{Y_m}{8m}\rho\nabla^2\rho$, although not present in the bare action, is generated during the
RG flow and produces a separation of the mass renormalization into distinct longitudinal ($Z_\sigma=Z_m+\rho_0Y_m$) and Goldstone  ($Z_\ctheta=Z_m$) renormalization
factors~\cite{gersdorff_nonperturbative_2001,jakubczyk_longitudinal_2017}. Additionally, a second order  time-derivative term has been neglected (more details are given in Subsection~\ref{sec:Broken,sub:Regimes}).

The function $U(\rho,\mu)$ is the effective potential expressed in terms of the density $\rho(x)=\phid(x)\phi(x)$. 
We expand this potential to quartic order in the fields around its $k$-dependent minimum $\rho_0=\langle\rho\rangle$, and to
first order around the $k$-independent physical chemical potential $\mu_0$ so we can extract the boson density \cite{floerchinger_functional_2008}
\begin{multline}
 U(\rho,\mu)=u_0+u_1(\rho-\rho_0)+\frac{u_2}{2}(\rho-\rho_0)^2-n_0(\mu-\mu_0)-n_1(\mu-\mu_0)(\rho-\rho_0)-\frac{n_2}{2}(\mu-\mu_0)(\rho-\rho_0)^2.
 \label{eq:Ueff}
\end{multline}
Here the coefficients $u_i$ and $n_i$ all run with $k$. As  will be explained in Sec.~\ref{sec:Thermo}, $n_0$ is
the $k$-dependent boson density, reaching its physical value at $k=0$. The truncation  (\ref{eq:Gammak}--\ref{eq:Ueff}) is in line with ones commonly used in Cartesian FRG treatments, see, e.g., Ref.~\cite{floerchinger_functional_2008}. Limitations of this choice will  be discussed in Sec.~\ref{sec:Results}.

In this work we focus on a Bose gas in its superfluid phase where $\rho_0>0$ and $u_1=0$ for \emph{all} $k$. 
In this case the quantity
\begin{equation}
\rho_s=Z_m \rho_0,
\end{equation}
corresponds to the $k$-dependent stiffness with respect to phase changes, 
which at zero temperature should be equal to the boson density $n_0$. 
The stiffness $\rho_s$ is usually an approximate expression for the superfluid density (see Refs.~\cite{popov_functional_1987,pitaevskii_bose-einstein_2016}).
This approximation is valid in three dimensions and in the weakly-interacting regime in two-dimensions, and so we will use
$\rho_s$ as an approximation for the superfluid density in these cases.
However, this is not the case in one dimension,  where the superfluid density can only be extracted directly from the free energy 
\cite{giamarchi_persistent_1995}. 
A detailed discussion of the difference between the stiffness $\rho_s$ and the superfluid density can be found in Ref.~\cite{prokofev_two_2000}.

\section{Broken phase}
\label{sec:Broken}

In this section we discuss the interpolating representation used for the fields  and the resulting flow equations.
We also give a summary of the most important aspects of the field representations and the interpolation scheme. 
For a more detailed discussion see Ref.~\cite{isaule_application_2018}.

\subsection{Field representations}

Because the effective potential has a non-zero minimum at $\rho_0=\langle\phid\phi\rangle$, we  
define the fluctuating boson fields relative to $\rho_0$. The most common decomposition is the
Cartesian representation, where the boson fields are parametrized as 
\begin{equation}
\phi=\sqrt{\rho_0}+\sigma+i\pi.
\label{eq:PhiCart}
 \end{equation}
Here $\sigma$ describes the longitudinal fluctuations and $\pi$ the fluctuations of the 
gapless Goldstone mode. An alternative parametrization  is the AP  representation
as introduced in the hydrodynamic effective theory \cite{popov_functional_1987}. In this work the AP representation is given in the form
\begin{equation}
\phi=(\sqrt{\rho_0}+\sigma)e^{i\ctheta/\sqrt{\rho_0}},
\label{eq:PhiAP}
\end{equation}
where $\sigma$ now describes  radial fluctuations and
$\ctheta$ fluctuations of the Goldstone (phase) mode.

Following Popov's approach \cite{popov_functional_1987}, we have proposed a $k$-dependent parametrization of the boson fields in Ref.~\cite{isaule_application_2018}.
This is constructed so that we use the AP representation in the IR and the Cartesian representation in the UV, with a smooth change of representation
between the two. This interpolating representation is given by \cite{lamprecht_confinement_2007},
\begin{equation}
\phi=(\sigma+\bk)e^{i\ctheta/\bk}-(\bk-\sqrt{\rho_0}).
\label{eq:InterpFields}
\end{equation}
Here the function $\bk$ must tend to $+\infty$ as $k\to\infty$  so it gives the Cartesian representation,
$\phi(\X) = (\sqrt{\rho_0}+\sigma(\X))+i\ctheta(\X)$,
while it must tend to $\sqrt{\rho_0}$ for $k\to 0$ where it gives the AP representation,  Eq.~(\ref{eq:PhiAP}).
By varying $\bk$ as $k$ runs, the fields smoothly change representation during the flow.
The resulting parametrization of the ansatz for the effective action and the flow equations are given in Appendix~\ref{app:Interpolating}.
(The specific form of the function $\bk$ used here is given below in Subsection \ref{sec:Broken,sub:Regimes}.)

As discussed in detail in Ref.~\cite{isaule_application_2018}, 
one important aspect of the use of Eq.~(\ref{eq:InterpFields})  is that there is a change of interpretation of $\rho_0$ with $k$. 
Whereas in the Cartesian representation $\rho_0$ corresponds to the scale-dependent condensate density $\rho_c$,
in the AP representation $\rho_0$ corresponds to the scale-dependent quasi-condensate density $\rho_q$ \cite{kagan_influence_1987},
which can be quite different from $\rho_c$.
As proven in Ref.~\cite{isaule_application_2018}, by using the interpolating representation $\rho_0$ correctly changes from
$\rho_c$ to $\rho_q$ during the flow.
It is this feature that enables us to obtain a finite $\rho_q$, and hence a finite stiffness $\rho_s$, when the system shows QLRO and $\rho_c=0$.

\subsection{Gaussian and Goldstone regimes}
\label{sec:Broken,sub:Regimes}

As  argued in Ref.~\cite{isaule_application_2018}, the Cartesian representation must be used in 
UV regime where both longitudinal and Goldstone fluctuations are important, that is, 
where the contribution $2u_2\rho_0$ is small compared to the kinetic term. In that regime the path integral over fluctuations is approximately Gaussian.
On the other hand, the AP representation needs to be used in the IR regime where the Goldstone mode dominates over the amplitude mode.
These two regimes can be distinguished in the FRG flow by the dimensionless quantity \cite{wetterich_functional_2008}
\begin{equation}
w_k=\frac{Z_\sigma k^2/2m}{2u_2\rho_0},
\end{equation}
where $Z_\sigma=Z_m+\rho_0 Y_m$. We refer to the regime where $w\gg 1$ as the Gaussian regime and 
the regime where $w\ll 1$ as the Goldstone regime. If the system has a finite physical stiffness, 
the flow starts from scales deep in the Gaussian regime,
and it ends in the Goldstone regime. The transition between the two regimes can be characterized by the scale $k_h$ where $w=1$,
which we refer to  as the \textit{healing scale}, in analogy to the physical healing length \cite{pitaevskii_bose-einstein_2016}.

We take the following form for the function $\bk$ \cite{isaule_application_2018,lamprecht_confinement_2007},
\begin{equation}
\bk=\sqrt{\rho_0}\left[1+(\alpha \, w_k)^\nu\right].
\label{eq:bk}
\end{equation}
This has the required behaviors in the limits $k\to\infty$ and $k\to 0$.
The parameter $\alpha$ controls the specific scale where the transition between representations is made and $\nu$ determines the rate of switching.
In our previous work we have concluded that at the level of truncation used in this work, reasonable choices of $\alpha$ lie between $0.5$ and $2.0$, and that for $\nu\geq 2.5$
the results converge. In the following, we shall use $\alpha=1$ and $\nu=3$.

Another important change in each regime is the form of the dispersion relation, which changes from
particle-like ($\epsilon_\Q=\Q^2/2m$) in the Gaussian regime to phonon-like ($\epsilon_\Q=c_s \Q$) in the Goldstone regime.
This is closely related to the onset of superfluidity \cite{andersen_theory_2004,posazhennikova__2006}.
At the level of truncation employed here, the microscopic sound velocity  obtained from the propagator
is given by \cite{dupuis_non-perturbative_2007,floerchinger_functional_2008},
\begin{equation}
c_s=\left(\frac{Z_m/2m}{Z_\phi^2/2u_2\rho_0}\right)^{1/2}\bigg|_{k=0}.
\label{eq:cs}
\end{equation}
We stress that other authors include a  term  of the form $V_\phi \phid\partial^2_\tau\phi$ in the action
as it is necessary to obtain a finite sound velocity when using the Cartesian representation ~\cite{dupuis_non-perturbative_2007,wetterich_functional_2008}.
As we show later, by using the AP representation in the Goldstone regime both $Z_\phi$ and $u_2$ saturate 
in the physical limit, so no second order term is required. 
We  note that this second-order coupling can nevertheless be generated during the flow, although
its inclusion is beyond the scope of this work.

One important property of the microscopic sound velocity $c_s$, is that at zero temperature
is equal to the macroscopic sound velocity $v_s$ of the system  \cite{gavoret_structure_1964}. The latter is related to
the macroscopic properties of the system through $v_s=(\partial P/\partial n_0)^{1/2}$, where $P$ is the pressure and $n_0$ the density.
We use this property to check that our interpolating approach gives $c_s$ correctly.
As we show in Sec.~\ref{sec:Results}, at $T=0$ we obtain a reasonable agreement between our results for $c_s$ and
known values of $v_s$.

\section{Thermodynamics}
\label{sec:Thermo}

The value of the fully evolved effective potential at the minimum $U(\rho_0,\mu_0)$ corresponds
to the density of the grand canonical potential $\Omega_G$. Since the differential of $\Omega_G$ is given by
\begin{equation}
d\Omega_G=-PdV-S dT-Nd\mu,
\end{equation}
the thermodynamic properties of the system can be extracted by taking derivatives of $U$.
Running quantities can be defined at any scale in the RG flow and take their physical
values at $k=0$. This leads us  to identify
\begin{equation}
n_0=-\frac{\partial U}{\partial \mu}\Big|_{\rho_0,\mu_0},
\end{equation}
as the scale-dependent boson density, and
\begin{equation}
s=-\frac{\partial U}{\partial T}\Big|_{\rho_0,\mu_0},
\end{equation}
as the scale-dependent entropy density. 
Then, in addition to flow equations listed in Eq.~(\ref{eq:FlowEqs_b}), we can follow the evolution of $s$ by using
\begin{equation}
\partial_k s=-\partial_T (\partial_k U)=\partial_T(\partial_k \Gamma)|_{\rho_0,\mu_0}.
\label{eq:floweq_s}
\end{equation}
In this method, we evaluate the derivative with respect to temperature in Eq.~(\ref{eq:floweq_s}) after performing the sums over Matsubara 
frequencies.

The scale-dependent pressure is given by $P=-u_0$, thus its evolution could in principle be solved directly from
$\partial_k u_0=\partial_k \Gamma|_{\rho_0,\mu_0}$ (see for example Refs.~\cite{rancon_thermodynamics_2012,rancon_universal_2012}).
However, as noted by Blaizot \emph{et al.} \cite{blaizot_perturbation_2007}, since the canonical dimension of $u_0$ is $[k^{d+2}]$, the renormalization of the vacuum pressure at $k=\Lambda$ 
requires several counter terms and the values of these can be difficult to determine within a numerical calculation. 
Instead, we compute the pressure from the boson density and entropy density, making the plausible and rather straightforward assumption
that the counter terms are independent of $T$ and $\mu_0$. 
From the Maxwell relations,
\begin{equation}
n_0=\frac{\partial P}{\partial \mu}\bigg|_{T,V}, \qquad s=\frac{\partial P}{\partial T}\bigg|_{\mu_0,V},
\end{equation}  
we see that the pressure can be obtained by integrating the physical values of $n_0$ and $s$ for a fixed temperature and chemical potential, respectively. This gives
\begin{align}
P(\mu_0,T)=&P(\mu_0=0,T)+\int_0^{\mu_0} n_0(\mu',T)d\mu', \label{eq:P_intmu}\\
P(\mu_0,T)=&P(\mu_0,T=0)+\int_0^T s(\mu_0,T')dT'.\label{eq:P_intT}
\end{align}
We know that in the vacuum limit $P(\mu_0=0,T=0)=0$. 
Starting from this value, we compute $P(\mu_0,T=0)$ using Eq.~(\ref{eq:P_intmu}), and from that we compute $P(\mu_0,T>0)$ with  Eq.~(\ref{eq:P_intT}).

Once we have determined the pressure, it is then possible to evaluate the energy density $\epsilon$ of the system. In the grand-canonical formalism this is given by
\begin{equation}
\epsilon=-P+n_0\mu_0+s T,
\end{equation}
and from that we can obtain the energy per particle, $E/N=\epsilon/n_0$.

We note that it has been argued that problems in following the FRG flow of the pressure may be caused by the use of frequency-independent regulators. These are thought to lead to an incorrect behavior of the Matsubara sums since both Matsubara frequencies larger and smaller than the cut-off scale contribute on an even footing \cite{floerchinger_nonperturbative_2009}.
Floerchinger and Wetterich address this in Ref.~\cite{floerchinger_nonperturbative_2009} by requiring the wave-function and mass renormalization factors to take bare values at high 
frequencies. Alternatively, 
Blaizot \emph{et al.} in Ref.~\cite{blaizot_calculation_2011} propose an alteration of the domain of the frequency integration in vacuum, so it matches
the domain of the Matsubara sums. 
Our approach, in contrast, makes no such modifications: it relies only on the independence of the counter terms on $T$ and $\mu_0$, and it is robust and applicable to both the normal
and superfluid phase of Bose gases. 
As we show in the next section, it gives good results for the energy per particle
as extracted from the pressure, in all cases studied, supporting this assertion.

\section{Initial conditions and renormalization of the interaction}
\label{sec:IC}

We start the flow at a scale $k=\Lambda$ that is much larger than the relevant scales of the problem ($k_h$ and $k_T=\sqrt{2\pi m T}$),
and where the RG flow is insensitive to many-body effects.
This can  easily be seen in the propagator (Eq.~(\ref{eq:Gb})), where, in the UV, the  term $2u_2 \rho_0$ in the longitudinal propagator
is small compared to the kinetic term, resulting in a flow that is similar to that in the
symmetric phase, where $\rho_0=0$. Also, thermal effects are small in the UV, resulting in a flow 
that behaves like that for zero temperature. As a result, the UV flow approaches that in vacuum ($T=0$, $\mu_0\leq 0$), which
can therefore be used to fix the initial conditions.

With the ansatz used for the effective action, Eqs.~\eqref{eq:Gammak} and \eqref{eq:Ueff},  only $u_2$ and $n_2$ run in vacuum, and thus all the other couplings
can be taken as their bare values,
\begin{align}
&\rho_0(\Lambda)=n_0(\Lambda)=\frac{\mu_0}{u_2(\Lambda)}\Theta(\mu_0),\quad u_1(\Lambda)=-\mu_0\Theta(-\mu_0), \nonumber \\
&Z_m(\Lambda)=Z_\phi(\Lambda)=1, \quad Y_m(\Lambda)=0, \quad n_1(\Lambda)=1, \quad s(\Lambda)=0.
\end{align}
In addition, although the coupling $n_2$ flows in vacuum, it vanishes in the UV, and thus  we can set $n_2(\Lambda)=0$ to a good approximation.
On the other hand, the interaction term $u_2$ needs to be renormalized as known from various RG approaches \cite{bijlsma_renormalization_1996}.
In vacuum, we can use the fact that 
$u_2$ at $k=0$ is related to the two-body $T$-matrix to connect the RG flow with physical scattering.

For a frequency-independent regulator, the flow of $u_2$ in vacuum is given by
\begin{equation}
\partial_k u_2 = \frac{u_2^2}{2}\int\frac{\ddQ}{(2\pi)^d} \frac{\partial_k R(\Q,k)}{(\Q^2/2m+|\mu|+R(\Q,k))^2},
\label{eq:vu2vacuum}
\end{equation}
which can be  solved in closed form to give
\begin{equation}
\frac{1}{u_{2,\Lambda}}-\frac{1}{u_{2,k}}=\frac{1}{2}\int\frac{\ddQ}{(2\pi)^d}\left[ \frac{1}{\Q^2/2m+|\mu|+R(\Q,\Lambda)}-\frac{1}{\Q^2/2m+|\mu|+R(\Q,k)}\right].
\label{eq:u2vacuum}
\end{equation}
The approach to renormalizing  $u_2$  differs in one, two and three dimensions, as described in the following sections.

\subsection{Three dimensions}
In three dimensions and 
at low energy, the two-body interaction can be characterized by the vacuum $T$-matrix,
\begin{equation}
T_{2B}=\frac{4\pi a_{3D}}{m},
\label{eq:T2B_3D}
\end{equation}
where $a_{3D}$ is the $s$-wave scattering length. We then impose on  Eq.~(\ref{eq:u2vacuum}) the condition 
that in the physical limit $u_{2}(k=0)=T_{2B}$.
Since $T_{2B}$ is energy-independent,
we can solve Eq.~(\ref{eq:u2vacuum}) using $\mu_0=0$. With the regulator~(\ref{eq:Rexp}), the integrals can be performed
analytically, giving
\begin{equation}
u^{\text{exp}}_{2,\Lambda}=\left[\frac{m}{4\pi a_{3D}}-\frac{m}{4\pi^{3/2}}\Lambda\right]^{-1},
\label{eq:u2Lambda_3D_Exp}
\end{equation}
An analogous initial condition is given by Floerchinger and Wetterich 
\cite{floerchinger_functional_2008} for an optimized regulator \cite{litim_optimized_2001}. 
If we require $u_{2}(\Lambda)$  to be finite and positive, we find that there is an upper bound for the value of the scattering length.
Thus, although $\Lambda$ must be chosen to be much larger than $k_h$ and $k_T$, it cannot be chosen to be arbitrarily large.
The constraint $u_{2}(\Lambda)>0$ actually ensures that we always work with a weakly-interacting Bose gas, where $n_0 a^{1/3}$ is small.

\subsection{Two dimensions}

In the case of a two-dimensional gas, the two-body $T$-matrix at low energies can be written as
\begin{equation}
T_{2B}(-2\mu)=\frac{4\pi/m}{\log(2/|\mu| m a_{2D}^2)-2\gamma_E},
\label{eq:T2B_2D}
\end{equation}
where $a_{2D}$ is the two-dimensional scattering length 
and $\gamma_E\approx 0.5772$ is Euler's constant. 
Here $-2\mu$ acts as the energy of the two-body system. 
In experiments, two-dimensional systems are usually achieved by confining a trapped three-dimensional gas into a
two-dimensional configuration. This allows $a_{2D}$ to be related to the original three-dimensional scattering length 
$a_{3D}$ and the size of the confinement $a_z=\sqrt{1/m\omega_z}$,
where $\omega_z$ is the frequency of the harmonic confinement potential. These parameters are then related by\cite{petrov_interatomic_2001}
\begin{equation}
a_{2D}=a_z \left(2\sqrt{\frac{\pi}{A}}e^{-\gamma_E}\right)\exp\left(-\sqrt{\frac{\pi}{2}}\frac{a_z}{a_{3D}}\right),
\end{equation} 
where $A\approx 0.91$. Since the length $a_z$ sets the scale that separates the two and three dimensional regimes, 
as long as we restrict the flow to values of $k$ less than   $\Lambda \ll a_z^{-1}$,  the system is effectively two-dimensional.

Unlike the three-dimensional version, this $T_{2B}$ 
remains dependent on the energy of the two-body system $-2\mu$, even at low energies.
Moreover, it vanishes for zero energy; a behavior that is correctly recovered by the flow of $u_2$ for $\mu_0=0$,
which vanishes as $\sim\log^{-1}(k/\Lambda)$.

As noted in Ref.~\cite{lim_correlation_2008}, in order to fix the initial condition for $u_2$ we have to consider a 
system in vacuum but at finite energy ($\mu_0<0$). The integral in Eq.~(\ref{eq:u2vacuum}) cannot be performed analytically
with the regulator~(\ref{eq:Rexp}). A workaround is to modify  the regulator to
\begin{equation}
R(\Q,k)=\frac{\Q^2/2m+|\mu_0|}{\exp\left(\frac{\Q^2+2m|\mu_0|}{k^2}\right)-1}.
\end{equation} 
This gives
\begin{equation}
u^{\text{exp}}_{2,\Lambda}=\left[\dfrac{1}{u_{2,0}}-\dfrac{m}{4\pi}\Gamma(0,2m|\mu_0|/\Lambda^2)\right]^{-1},
\end{equation}
where $\Gamma(0,x)$ is the incomplete Gamma function. By replacing $u_{2,0}=T_{2B}$ and then taking the limit $\mu_0\to 0$ we obtain,
\begin{equation}
u^{\text{exp}}_{2,\Lambda}=\frac{4\pi}{m}\left[-\gamma_E-\log\left(a_{2D}^2 \Lambda^2/4\right)\right]^{-1},
\label{eq:u2Lambda_2D_Exp}
\end{equation}
which we use as the initial condition in two dimensions. We note that Lammers \emph{et al.} \cite{lammers_dimensional_2016} obtained
an analogous initial condition using an optimized regulator.
We stress that although we take the limit $\mu_0\to 0$, the
initial condition is still valid for finite $\mu_0$, as the presence of a finite chemical potential in the flow equations
is enough to capture the many-body effects during the flow \cite{lim_correlation_2008}. Indeed, we have checked that, using initial condition (\ref{eq:u2Lambda_2D_Exp}), 
we recover  $u_2=T_{2B}(-2\mu_0)$ at $k=0$.

We also stress that this procedure to obtain the initial condition is valid for any regulator where the logarithmic energy terms cancel
and thus the limit $\mu_0\to 0$ is well defined. As in three-dimensions,  the initial condition constrains the value of $a_{2D}$, 
again restricting the system to a weakly-interacting Bose gas where $\log^{-1}(1/n_0a_{2D}^2)$ is small.

\subsection{One dimension}

The one-dimensional gas is significantly different from its two and three dimensional counterparts, since this problem is dominated by IR fluctuations \cite{wetterich_functional_2008}.
The strength of the interaction is best characterized by the bare coupling \cite{olshanii_atomic_1998},
\begin{equation}
g_{1D}=-\frac{2}{m a_{1D}},
\end{equation}
where $a_{1D}$ is defined to be the scattering length in one dimension. 
Note that, for a repulsive interaction ($g_{1D}>0$), we should take $a_{1D}<0$. As in the two-dimensional case,
one-dimensional systems can be achieved experimentally by confining a three-dimensional gas, making
possible to relate $a_{1D}$ to $a_{3D}$ and the confinement length $a_\perp=\sqrt{1/m\omega_\perp}$.
These parameters are related by \cite{olshanii_atomic_1998}
\begin{equation}
a_{1D}=-\frac{a_\perp^2}{a_{3D}}\left(1+C\frac{a_{3D}}{a_\perp}\right).
\end{equation}
where $C=\zeta(1/2)/\sqrt{2}$. Again, as long as we restrict the flow to  $k\leq \Lambda\ll a_\perp^{-1}$, the system is effectively one-dimensional.

In vacuum $u_2$ vanishes linearly with $k$ as $k\to 0$ as a consequence of the dominance of IR fluctuations. In contrast, 
$u_2$ flows to a constant value in the UV.  
Thus, in one dimension, $u_2$ does not need to be renormalized, and its initial condition is simply  given by
\begin{equation}
u_{2,\Lambda}=-\frac{2}{m a_{1D}}.
\end{equation}
One drawback of this initial condition is that it does not force the system to be in the weakly-interacting regime.
As we discuss in  Appendix~\ref{app:TGregime}, as the strongly interacting regime is approached by decreasing the boson density, the flow starts to show incorrect behavior and becomes unstable.

\section{Results}
\label{sec:Results}

In the following, we present results for three, two and one dimensions. For each case we first compare the flows
obtained for the interpolating and Cartesian representations. 
From now on, we will refer to the mass renormalization $Z_m$ as $Z_\ctheta$, to emphasize that it describes
the renormalization of the Goldstone (phase) mode \cite{isaule_application_2018}.
Subsequently, we present results for thermodynamic properties and compare them with known results.

The two representations differ in the Goldstone regime, where the interpolating representation 
is in its AP limit. Moreover, with the interpolating representation, $Z_\ctheta$ changes from values greater than one in the Gaussian regime, to smaller than one in the Goldstone regime.
This reflects the fact that $\rho_0$ changes from a scale-dependent condensate density to a quasi-condensate density.
Similar behavior was seen in the classical system studied in Ref.~\cite{isaule_application_2018}, and a detailed discussion can be found there.
Here, we focus on features that are particular to dynamical Bose gases.

\subsection{Three dimensions}

\begin{figure*}
	\centering
    \subfloat{\includegraphics[width=0.45\textwidth]{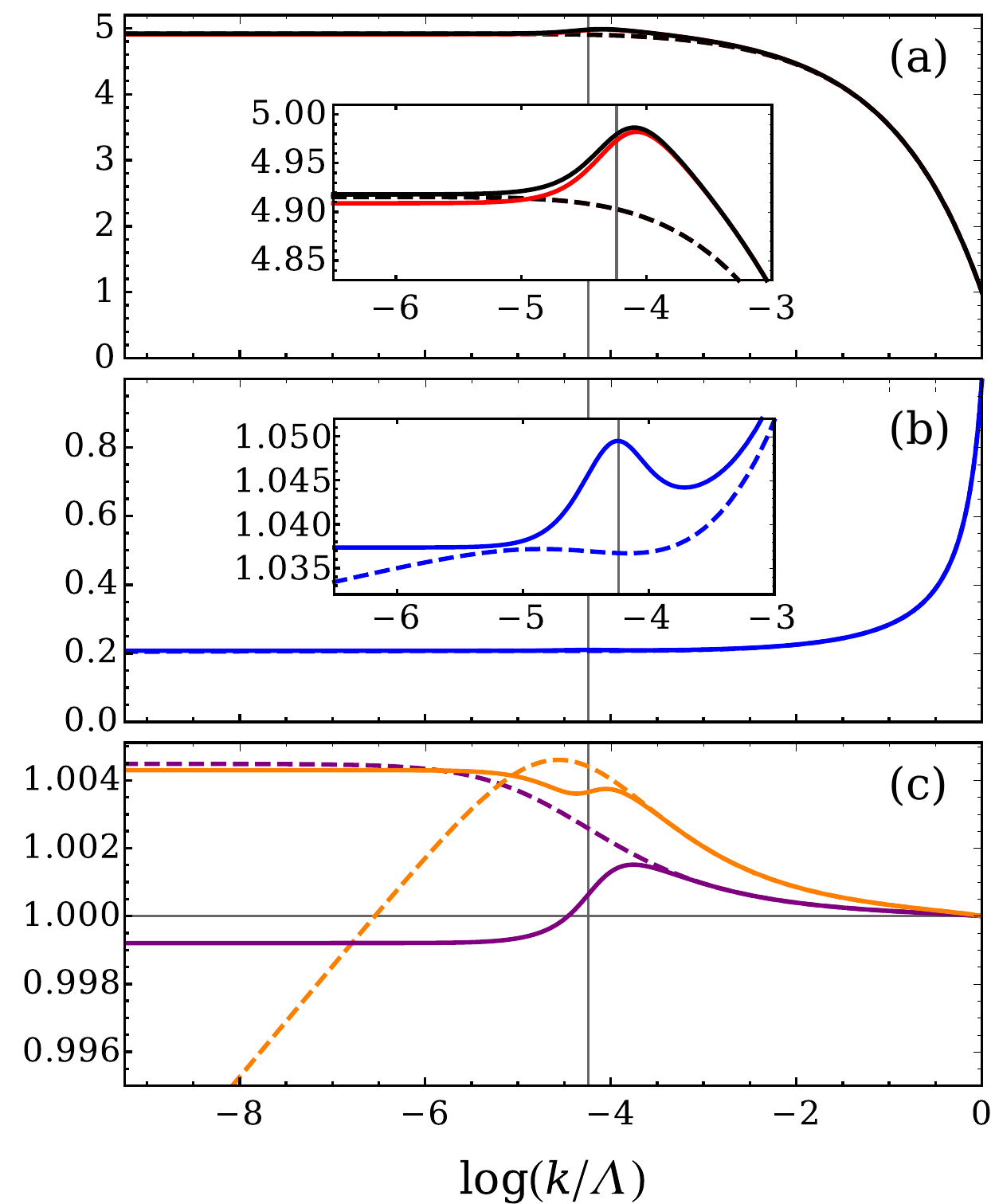}}
    \hspace{0.5cm}
    \subfloat{\includegraphics[width=0.45\textwidth]{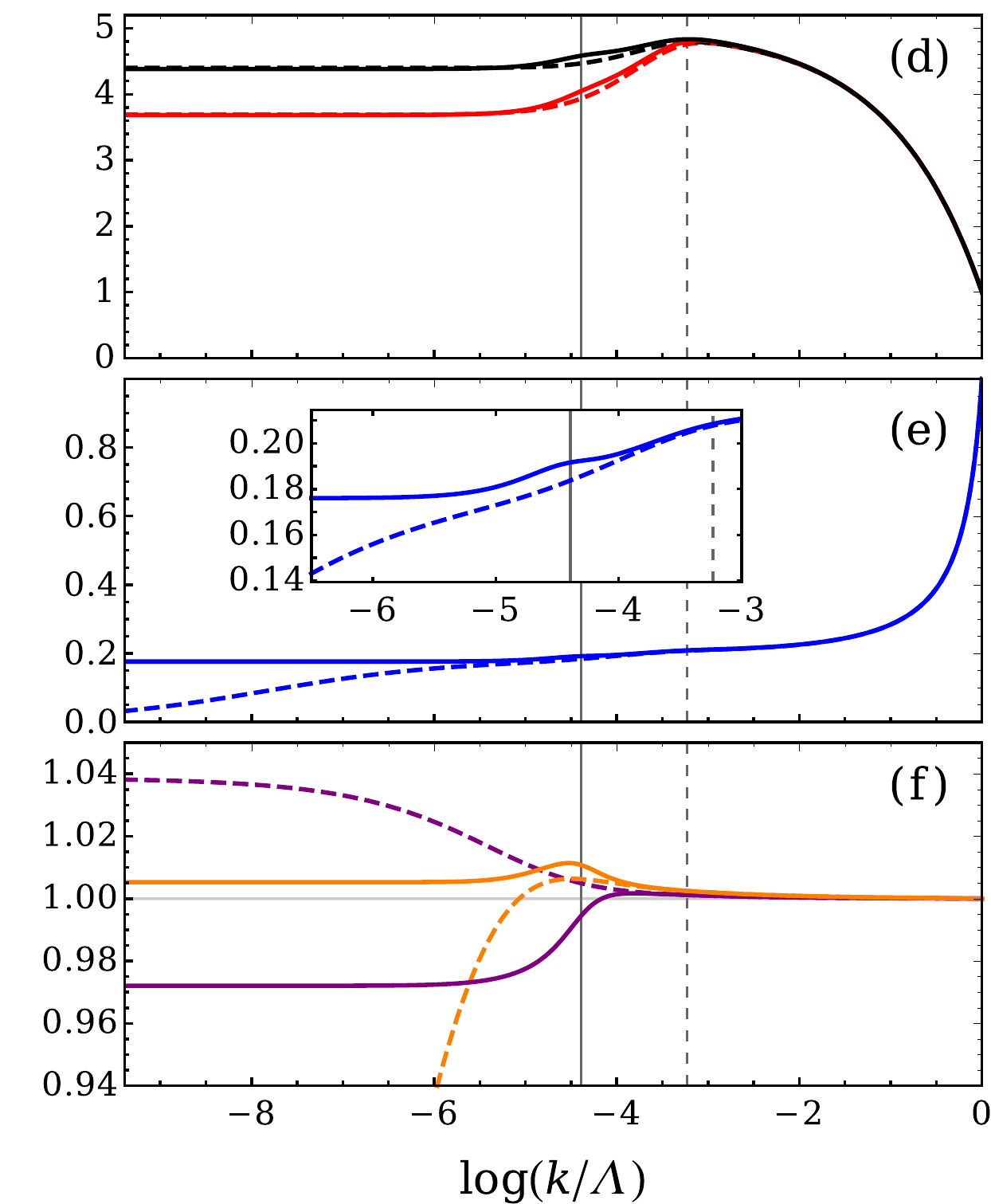}}
    \caption{Flow of $n_0$ (black lines, a and d), $\rho_s$ (red lines, a and d), $u_2$ (blue lines, b and e), $Z_\ctheta$ (purple lines, c and f)
    and $Z_\phi$ (orange lines, c and f) in three dimensions as functions of $\log(k/\Lambda)$ for $\mu_0=10^{-4}$. 
	(a,b,c) correspond to $T=0$ and (d,e,f) to $T=2\times 10^{-3}$. Both $\mu_0$ and $T$ are in units of $(m a_{3D}^2)^{-1}$. 
	The densities and $u_2$ have been rescaled by their values at the UV scale $\Lambda$.
    The solid lines are flows obtained using the interpolating representation, while the dashed lines  using the Cartesian representation.
    The vertical solid lines denote the healing scale $k_h$ and the vertical dashed lines the thermal scale $k_T=\sqrt{2\pi m T}$.
    The insets show details of the flows around $k_h$.} 
    \label{fig:Flows3D}   
\end{figure*}

Fig.~\ref{fig:Flows3D} shows some examples of flows in three dimensions at zero and finite temperature.
In the UV, $u_2$ decays as $k^{-1}$, as in vacuum, while the renormalization factors $Z_\phi$ and $Z_\ctheta$ do not run,
consistently with the boundary conditions. Using the Cartesian representation, both $u_2$ and $Z_\phi$ vanish in the physical limit as $(\log(k))^{-1}$
for $T=0$, and as $k$ for $T>0$. Since we do not include a quadratic time derivative term, this leads to a divergent microscopic sound velocity $c_s$.  
Note that the logarithmic approach to zero of $u_2$ is extremely slow and  is only visible when enlarged (see inset). 
On the other hand, for the interpolating representation, both $Z_\phi$ and $u_2$ quickly saturate to finite values, giving a finite $c_s$.

The boson and superfluid densities are rather insensitive to the choice of representation. 
At $T=0$,  with the Cartesian representation, we find $n_0=\rho_s$ during the flow, resulting in a completely superfluid system,
as expected. However, with the interpolating representation, the boson density becomes slightly greater than $\rho_s$.
This is likely to be an artifact of the truncation scheme.
We have checked that this difference is always smaller that 0.5 \%. 

\begin{figure}
	\centering
    \includegraphics[width=0.45\textwidth]{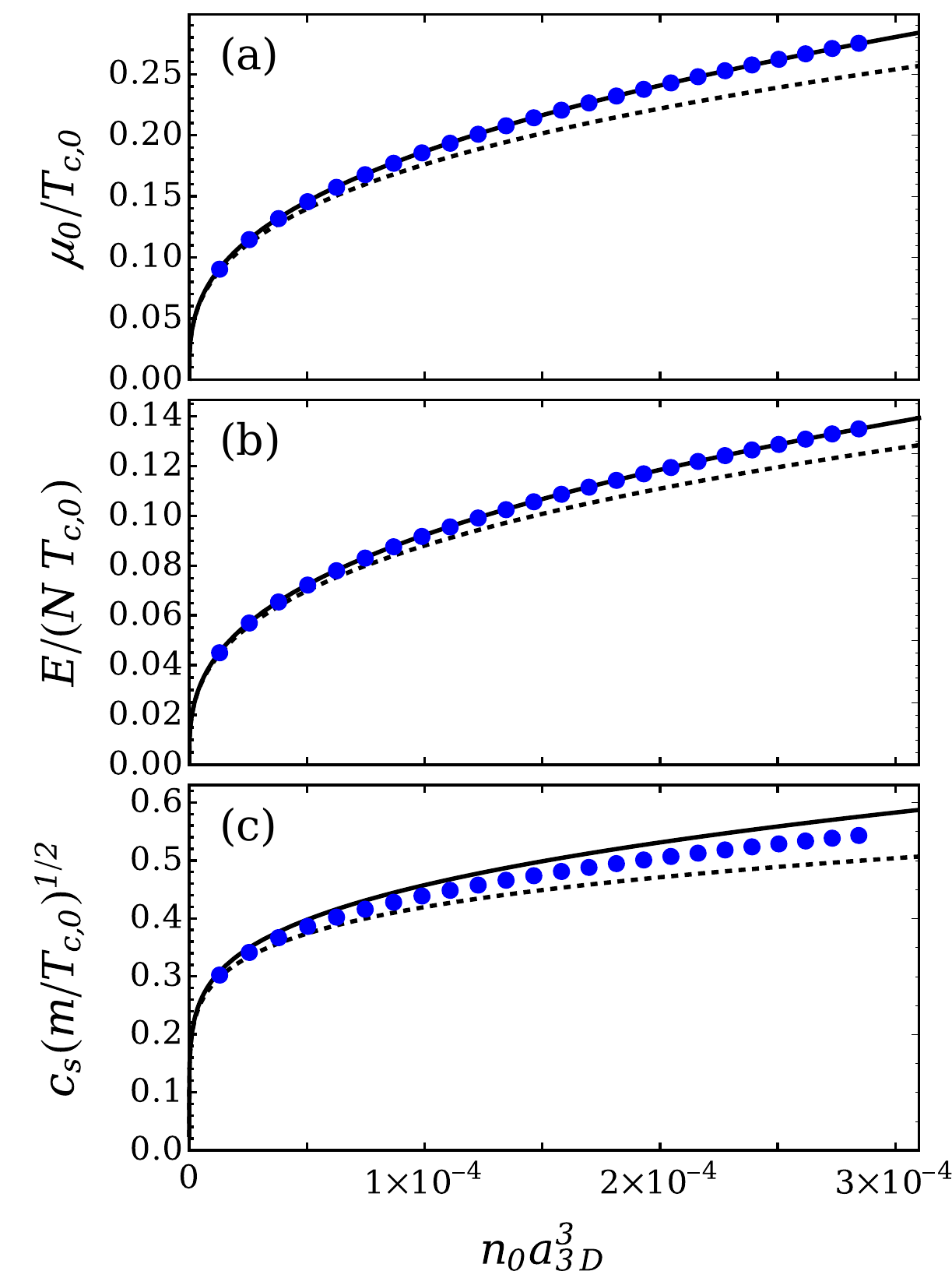}
    \caption{Chemical potential $\mu_0$ (a), energy per particle $E/N$ (b) and microscopic sound velocity $c_s$ (c) as 
    functions of  $n_0 a_{3D}^3$  in three dimension  at $T=0$. The blue circles are obtained using the interpolating representation. The dotted lines are the mean-field results, while the solid lines include the LHY corrections (\ref{eq:ZT_3D_mu}-\ref{eq:ZT_3D_vs}).} 
    \label{fig:ZT_Thermo_3D}   
\end{figure}

Fig. \ref{fig:ZT_Thermo_3D} shows results for the chemical potential, energy per particle and microscopic sound velocity at zero temperature
using the interpolating representation.
We work with dimensionless quantities written in terms of the critical temperature
of the ideal Bose gas,
\begin{equation}
T_{c,0}=\frac{2\pi}{m}\left(\frac{n_0}{\zeta(3/2)}\right)^{2/3}.
\end{equation}
We compare our results with the low-density expansions~\cite{lee_many-body_1957,lee_eigenvalues_1957}
\begin{align}
\mu_0\simeq &\frac{4\pi a_{3D} n_0}{m}\left(1+\frac{32}{3\sqrt{\pi}}(n_0a_{3D}^3)^{1/2}\right), \label{eq:ZT_3D_mu}\\
E/N\simeq &\frac{2\pi a_{3D} n_0}{m}\left(1+\frac{128}{15\sqrt{\pi}}(n_0a_{3D}^3)^{1/2}\right), \label{eq:ZT_3D_EoN}\\
v_s\simeq &\sqrt{\frac{4\pi a_{3D} n_0}{m^2}}\left(1+\frac{16}{\sqrt{\pi}}(n_0a_{3D}^3)^{1/2}\right), \label{eq:ZT_3D_vs}
\end{align}
where $v_s$ is the macroscopic sound velocity. The first terms inside the parentheses are the mean-field expressions,
whereas the second terms are known as the Lee-Huang-Yang (LHY) corrections \cite{lee_many-body_1957,lee_eigenvalues_1957}.
These  are based on a expansion in the small parameter $(n_0a_{3D}^3)^{1/2}$.
We stress that higher-order corrections depend on the details of the inter-particle potential.
For instance, the correction at order $n_0a_{3D}^3$ was first calculated in Ref.~\cite{braaten_quantum_1999}, where it was computed in terms of
the scale set by a van der Waals potential.
Since we use a contact interaction that depends only on the scattering length, we choose to not compare with those corrections.

The results for $\mu_0$ and $E/N$ show that the FRG correctly follows the LHY results. This is not surprising, as several works have
shown previously that the FRG can successfully describe these bulk thermodynamic properties \cite{floerchinger_nonperturbative_2009,rancon_thermodynamics_2012}.
The results for the microscopic sound velocity $c_s$ lie between the MF and LHY results for the macroscopic sound velocity $v_s$.
Here we remind the reader that at zero temperature both velocities should be equal \cite{gavoret_structure_1964}.
Our results show that the interpolating representation gives  reasonable results for $c_s$ at low densities, but
deviate at higher densities. This behavior is to be expected from the truncation employed in this work. For example, as mentioned previously, a term of second order in the energy should be
generated during the flow, and its inclusion will alter the physical value of $c_s$. 

\begin{figure}
	\centering
    \includegraphics[width=0.45\textwidth]{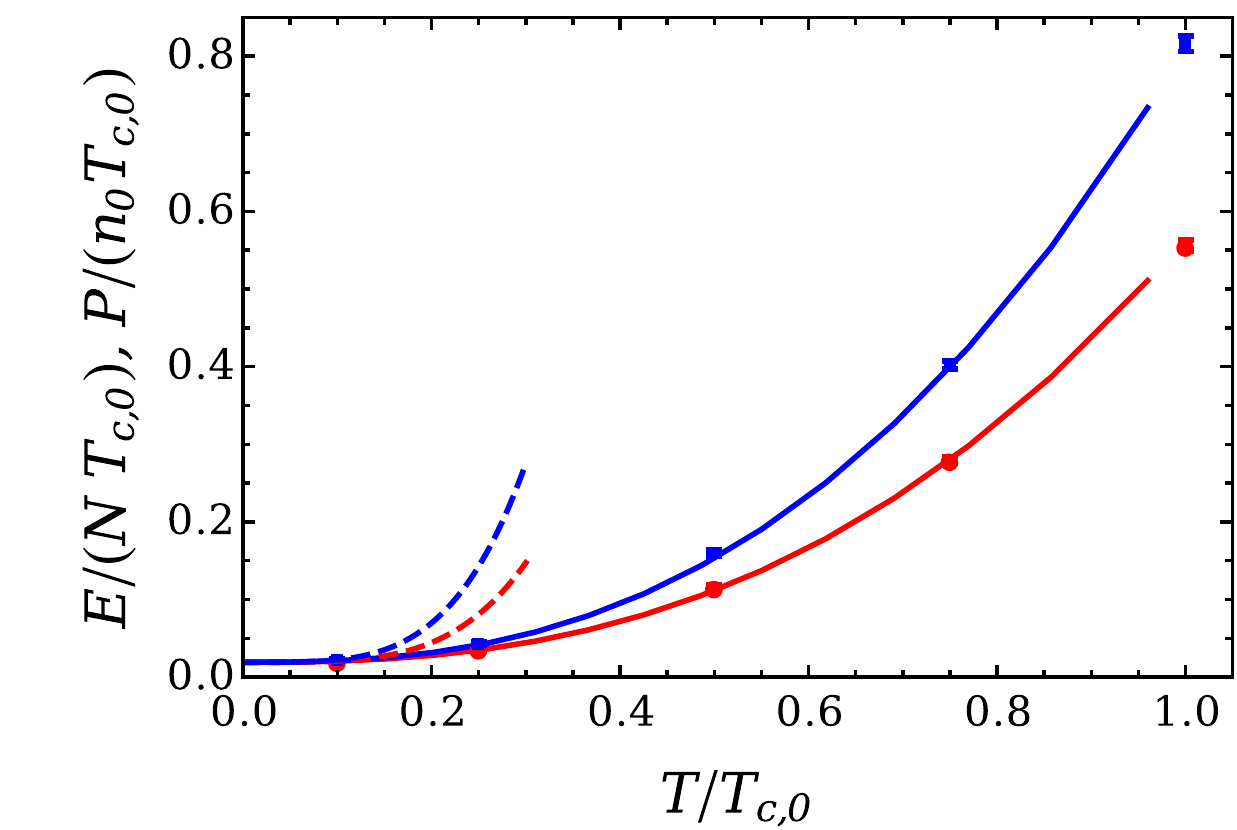}
    \caption{Energy per particle $E/N$ (blue) and pressure over density $P/n_0$ (red) as functions of $T/T_{c,0}$ for
    $n_0a_{3D}^3=10^{-6}$. The circles  are MC results from Ref.~\cite{pilati_equation_2006}, the dashed lines are mean-field 
    results~\cite{pitaevskii_bose-einstein_2016,andersen_theory_2004} and
    the solid lines are  results obtained using the interpolating representation.} 
    \label{fig:FT_Thermo_3D}   
\end{figure}

Fig. \ref{fig:FT_Thermo_3D} shows the energy per particle and pressure for  $n_0a_{3D}^3\approx10^{-6}$ at different temperatures.
These results are obtained from interpolating the FRG results at nearby densities, as the boson density flows with $k$ in our calculation.
We do no show results around the phase transition because the interpolation becomes unreliable at those temperatures.
We compare our results with Monte-Carlo simulations from Ref.~\cite{pilati_equation_2006} and with mean-field results at low 
temperatures (see Refs.~\cite{pitaevskii_bose-einstein_2016,andersen_theory_2004} for details).
The mean-field results are substantially less accurate for higher temperatures. In contrast, we obtain a 
reasonable agreement between our results and the simulations for both quantities.
Thus, from our results at zero and finite temperature we can confirm that our approach outlined in Sec.~\ref{sec:Thermo} correctly gives both the pressure
and the entropy.

\subsection{Two dimensions}

\begin{figure*}
	\centering
    \subfloat{\includegraphics[width=0.45\textwidth]{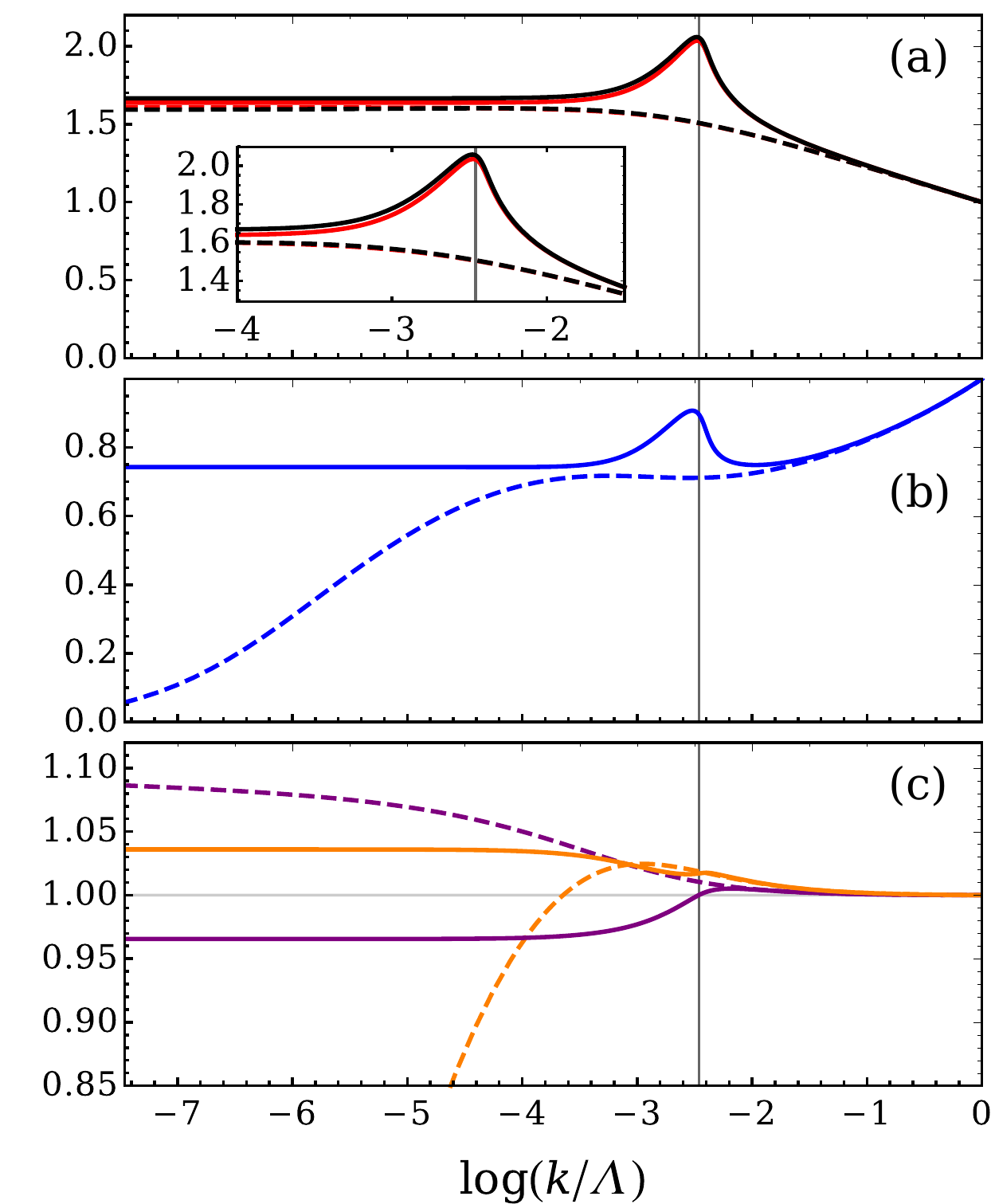}}
    \hspace{0.5cm}
    \subfloat{\includegraphics[width=0.45\textwidth]{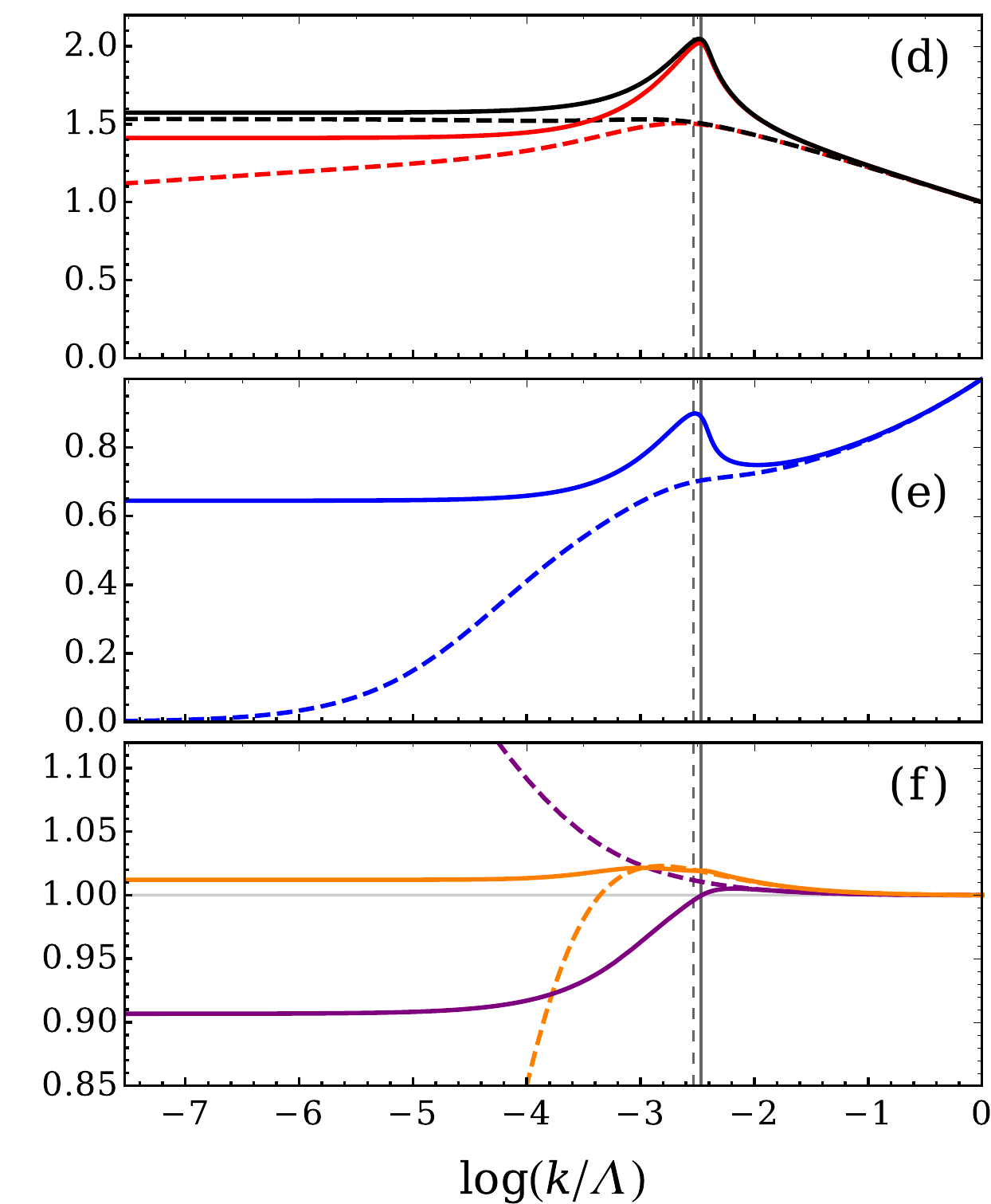}}
    \caption{Flow of $n_0$ (black lines, a and d), $\rho_s$ (red lines, a and d), $u_2$ (blue lines, b and e), $Z_\ctheta$ (purple lines, c and f)
    and $Z_\phi$ (orange lines, c and f) in two dimensions as functions of $\log(k/\Lambda)$ for $\mu_0=3.4\times 10^{-7} $. 
    (a,b,c) correspond to $T=0$ and (d,e,f) to $T=3.4\times 10^{-7}$. Both $\mu_0$ and $T$ are in units of $(m a_{2D}^2)^{-1}$. 
    The densities and $u_2$ have been rescaled by their values at the UV scale $\Lambda$.
    The solid lines are flows obtained using the interpolating representation, while the dashed lines using the Cartesian representation.
    The vertical solid lines denote the healing scale $k_h$ and the vertical dashed lines the thermal scale $k_T=\sqrt{2\pi m T}$.
    The inset (a) shows details of the flow of $n_0$ and $\rho_s$ at $T=0$ around $k_h$.} 
    \label{fig:Flows2D}      
\end{figure*}

In  Fig.~\ref{fig:Flows2D} we show some typical examples of flows   in the superfluid phase, both  at zero and finite temperature.
At finite temperatures, the two-dimensional system shows QLRO with a vanishing condensate density $\rho_c$ but a finite $\rho_s$.
As discussed above, this behavior cannot  be reproduced in the Cartesian representation under the truncations used to  date \cite{berges_non-perturbative_2002}
unless the regulator is fine-tuned. 
Indeed, as shown in Fig.~\ref{fig:Flows2D}d,
in the Cartesian representation, the superfluid density exhibits a slow decay in the Goldstone regime, which results in the flow reaching the symmetric phase
at a finite scale $k$, and hence in a non-superfluid system at $k=0$.
On the other hand, the interpolating representation corrects this by generating  a finite quasi-condensate density $\rho_q$ as well as
a finite superfluid density $\rho_s$ at $k=0$.

Overall, the flows display similar features to the three-dimensional case.
In the UV, $u_2$ shows the expected logarithmic vacuum-like behavior,  with $Z_\phi$ and $Z_\ctheta$ remaining at their
bare values.  At zero temperature in the IR, both $u_2$ and $Z_\phi$ vanish in the Cartesian representation, in this case linearly with $k$, again resulting in a diverging $c_s$
in the absence of quadratic time-derivative term. 
With the interpolating representation these quantities saturate at both zero and finite temperature, giving a finite $c_s$.
There is also a small depletion of the superfluid at zero temperature with the interpolating representation, which we have checked is
always less than 2.5\%. 

\begin{figure}
	\centering
    \includegraphics[width=0.45\textwidth]{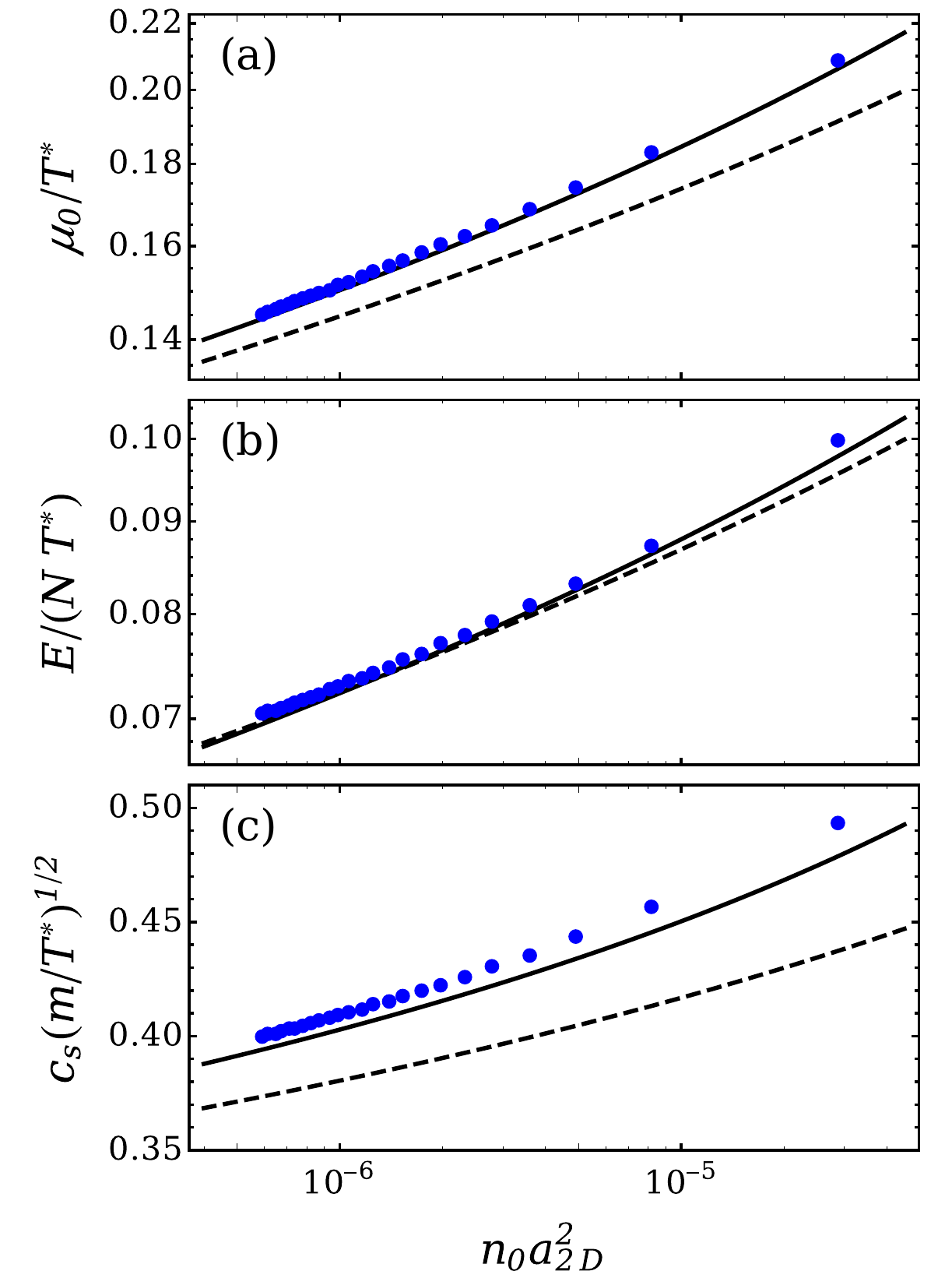}
    \caption{   Chemical potential $\mu_0$ (a), energy per particle $E/N$ (b) and 
    microscopic sound velocity $c_s$ (c) as functions of $n_0 a_{2D}^2$
    in two dimensions  at $T=0$. The blue circles are obtained using the interpolating representation. 
    The dotted lines are the Beliaev results at leading order, while the solid lines are the complete parametrization (\ref{eq:ZT_2D_mu}-\ref{eq:ZT_2D_cs}).} 
    \label{fig:ZT_Thermo_2D}   
\end{figure}

Fig.~\ref{fig:ZT_Thermo_2D} shows results for the chemical potential, energy per particle and microscopic sound velocity at zero temperature
with the interpolating representation, expressed in terms of the characteristic temperature $T^*=2\pi n_0/m$.
We compare our results with parametrizations for $\mu_0$ and $E/N$ obtained from MC simulations by Astrakharchik \emph{et al.} \cite{astrakharchik_equation_2009},
\begin{align}
\mu_0=&\frac{4\pi n_0/m}{|\log(x)|+\log|\log(x)|+C_1+\frac{\log|\log(x)|+C_2}{|\log(x)|}} \label{eq:ZT_2D_mu}\\
E/N=&\frac{2\pi n_0/m}{|\log(x)|+\log|\log(x)|+C_1+\frac{1}{2}+\frac{\log|\log(x)|+C_2+1/4}{|\log(x)|}}, \label{eq:ZT_2D_EoN}
\end{align}
where $x=n_0a_{2D}^2$, $C_1=-\log(\pi)-2\gamma_E-1$ and $C_2=-\log(\pi)-2\gamma_E+2$.
This parametrization is based on an expansion on the small parameter of the two-dimensional system $\log^{-1}(n_0 a_{2D}^2)$. 
At the lowest order, that is neglecting all the terms in the denominator except the  term $|\log(x)|$,
Eqs.~(\ref{eq:ZT_2D_mu}) and (\ref{eq:ZT_2D_EoN}) correspond to the expressions originally obtained by Schick \cite{schick_two-dimensional_1971} 
using the Beliaev method at leading-order, which can also be obtained from Popov's approach at tree level. Thus, we consider the expressions
at the lowest order as the MF level, and the complete forms as the higher-order corrections.

For the microscopic velocity we compare with the expression obtained by Schick \cite{schick_two-dimensional_1971} ,
\begin{equation}
v_s=\sqrt{\frac{4\pi n_0}{m^2|\log(na_{2D}^2)|}}, \label{eq:ZT_2D_cs}
\end{equation}
which again we will consider as the MF result. In order to compare with a higher-order estimate of the macroscopic sound velocity, we use parametrization 
Eq.~(\ref{eq:ZT_2D_mu}) and obtain the sound velocity  through $v_s=\sqrt{n/m\, \partial_n \mu_0}$.
The results of Fig.~\ref{fig:ZT_Thermo_2D}  show that the FRG  follows the parametrizations \eqref{eq:ZT_2D_mu} and \eqref{eq:ZT_2D_EoN} with  small deviations. 
As in three dimensions, we thus conclude that the FRG is able to correctly describe effects beyond MF.

\begin{figure}
	\centering
    \includegraphics[width=0.45\textwidth]{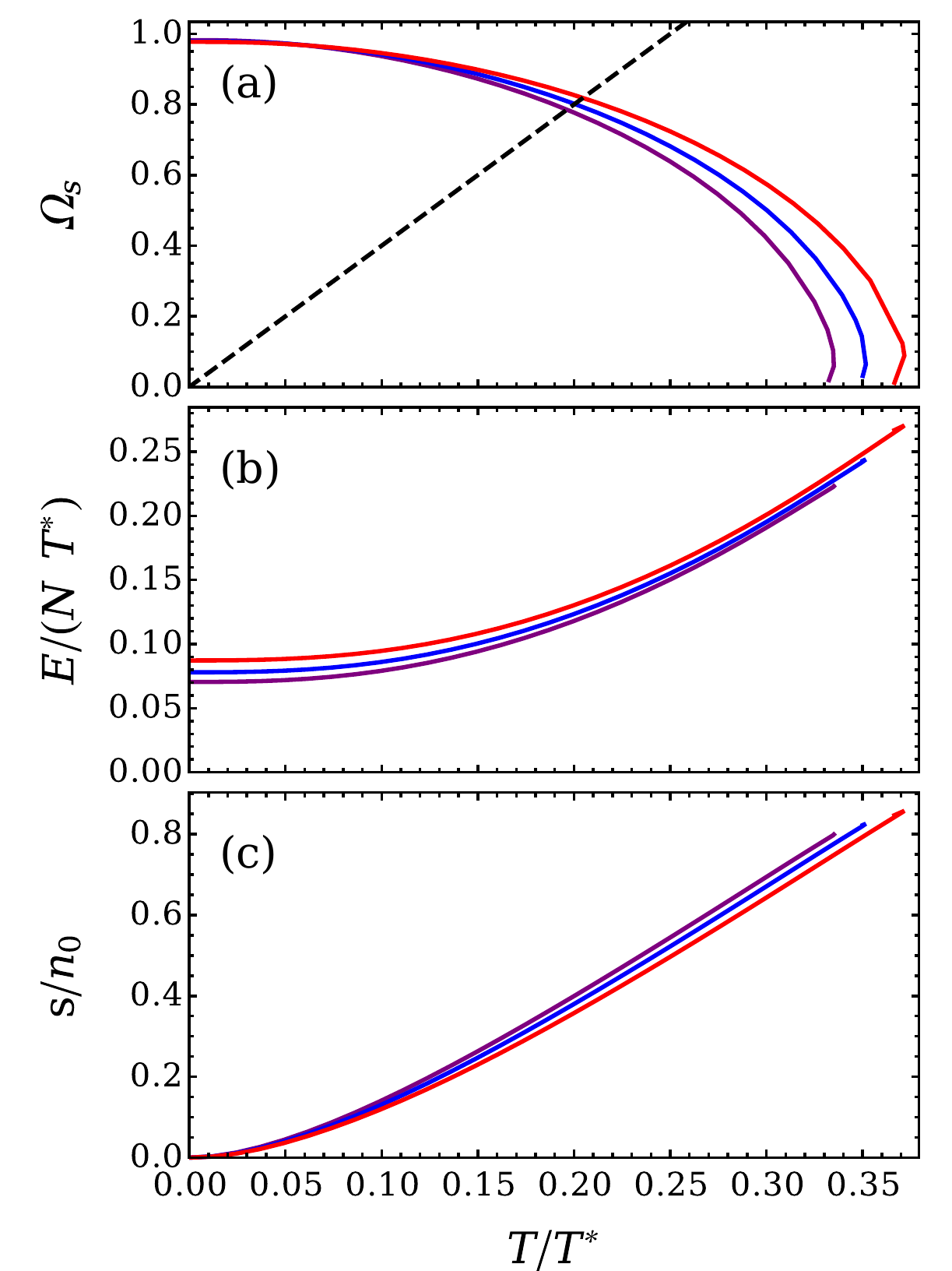}
    \caption{Superfluid fraction $\Omega_s$ (a), energy per particle $E/N$ (b) and entropy over density $s/n_0$ (c) 
    in two dimensions as functions of $T/T^*$ using the interpolating representation. 
	The black dashed line in (a) corresponds to the BTK critical superfluid density $\rho_{s,BKT}=2mT/\pi$.
	The purple curves are obtained for $\mu_0=5.1\times 10^{-7}$, the blue curves for $\mu_0=2.4\times 10^{-6}$ and the red curves for $\mu_0=3.7\times 10^{-5}$,
	with $\mu_0$ in units of $(ma_{2D}^2)^{-1}$.} 
    \label{fig:FT_Thermo_2D}   
\end{figure}

Fig.~\ref{fig:FT_Thermo_2D} shows the superfluid fraction, energy per particle and entropy over density at finite temperature for various
chemical potentials. As in our previous work, the superfluid density  goes smoothly to zero.  
However, in two dimensions the superfluid phase transition is driven by the unbinding of vortex pairs 
through the  Berenzinskii-Kosterlitz-Thouless (BKT) mechanism  \cite{berezinskii_v._l._destruction_1970,kosterlitz_ordering_1973}, where
the superfluid density shows a sudden jump from $\rho_{s,BKT}=2mT/\pi$ to zero instead of a smooth decrease.
Moreover, we encounter numerical instabilities in the region around the phase transition ($T/T^*\approx 0.35$ in the figure), which
are caused by an unphysical discontinuity shown by the boson density (see Ref.~\cite{isaule_application_2018} for a complete discussion).
Thus our results for $\rho_s\lesssim \rho_{s,BKT}$ are not realistic, due to the absence of vortices in our treatment which are
important at those temperatures.
In contrast, calculations using the Cartesian representation seem to be able to give a better description of the thermodynamic quantities
even though $\rho_s$ incorrectly vanishes in the superfluid phase \cite{rancon_universal_2012}.
The effect of the missing vortex physics on the critical temperature is discussed in Appendix~\ref{app:Tc2D}.

Our current calculation is  accurate only at $T\ll T_{BKT}$ where vortex effects are not important.
Indeed, we see that the  energy per particle and entropy show the expected increases with the temperature, and the superfluid fraction
shows a decrease in line with what can be found elsewhere \cite{pilati_critical_2008}.

\subsection{One dimension}

\begin{figure}
	\centering
    \includegraphics[width=0.45\textwidth]{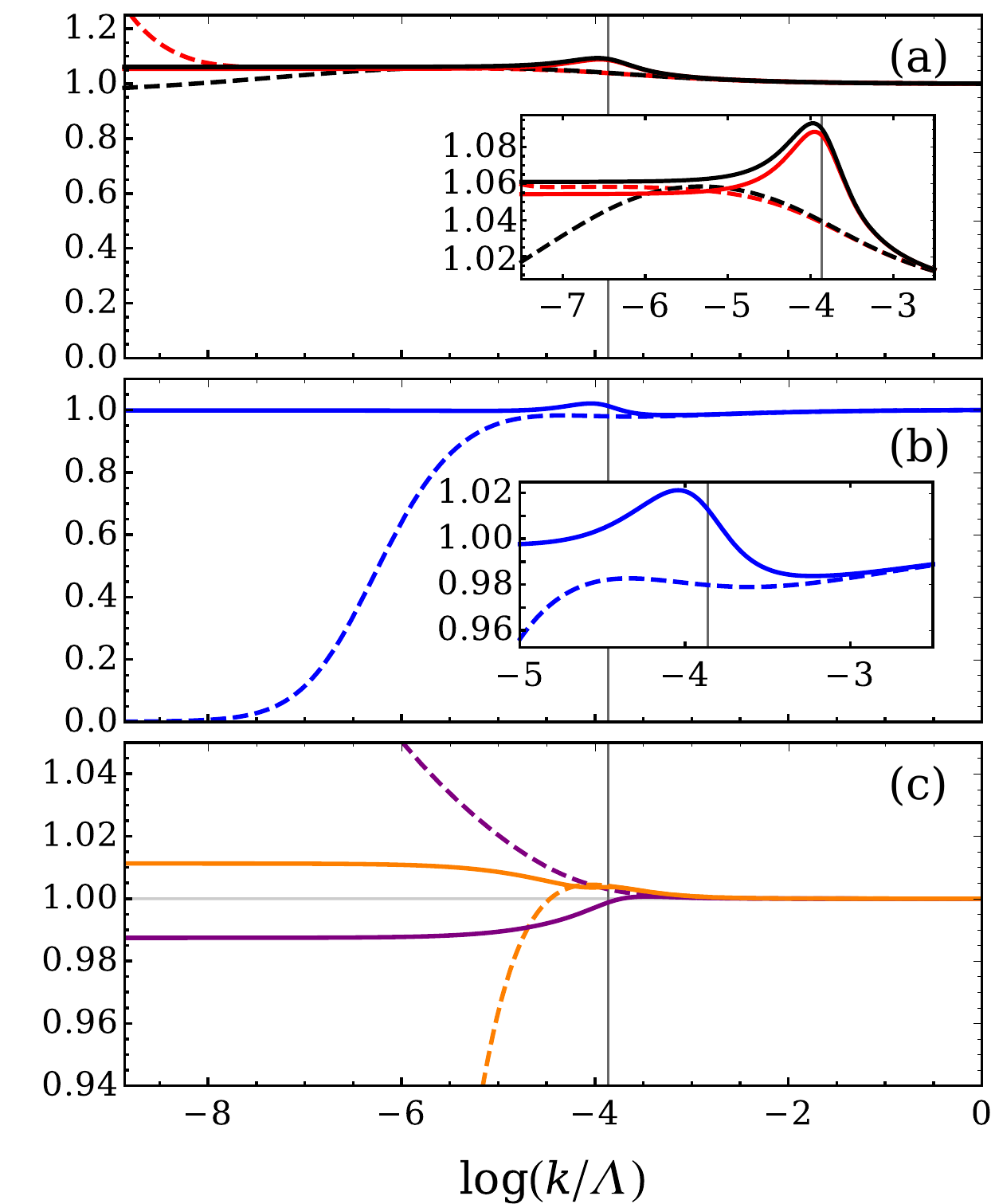} 
    \caption{Flow of $n_0$ (black lines, a), $\rho_s$ (red lines, a), $u_2$ (blue lines, b), $Z_\ctheta$ (purple lines, c)
    and $Z_\phi$ (orange lines, c) in one dimension at $T=0$ as functions of $\log(k/\Lambda)$ for $\mu_0=10^2$, with $\mu_0$ in units of $(m a_{1D}^2)^{-1}$. 
   The densities and $u_2$ have been rescaled by their values at the UV scale $\Lambda$.
    The solid lines are flows obtained using the interpolating representation, while the dashed lines using the Cartesian representation.
    The vertical solid lines denote the healing scale $k_h$.
    The insets show details of the flows around $k_h$.} 
    \label{fig:Flows1D}   
\end{figure}

In Fig.~\ref{fig:Flows1D} we show flows in one dimension at zero temperature in the weakly-interacting regime.
We see that the parameters in the flow are constant in the Gaussian regime. As discussed in Sec.~\ref{sec:IC}, $u_2$ should not flow in the UV in one dimension,
and thus our results are consistent with this behavior at high scales. Since the one-dimensional Bose gas at zero temperature shows QLRO, it shares features with the two-dimensional gas at finite temperatures.
As in that case, with the Cartesian representation the flow reaches the symmetric phase at a finite scale $k$, resulting in an incorrect normal 
phase, whereas with the interpolating representation we correctly obtain a finite $\rho_s$ in the physical limit.
Similarly, the interpolating representation enables us to obtain a finite sound velocity $c_s$ at the used order of truncation by giving a finite $u_2$ and $Z_\phi$ at $k=0$.

As in higher dimensions, we find that $\rho_s<n_0$ with the interpolating representation, but the difference between these quantities is less than $5 \%$ in the weakly-interacting regime studied here.
This difference becomes larger if we approach the strongly-interacting regime, as we discuss in Appendix~\ref{app:TGregime}.
We also note that, while the Cartesian representation can be used in two dimensions to extract some properties of the system, 
this is not the case in one dimension, where the gradient expansion quickly fails when the anomalous dimensions become large \cite{dupuis_non-perturbative_2007}.

The one-dimensional Bose gas is characterized by the dimensionless parameter
\begin{equation}
\gamma=-\frac{2}{n_0 a_{1D}}.
\end{equation}
The weakly-interacting MF regime corresponds to $\gamma\ll 1$, whereas the strongly-interacting Tonks-Girardeau (TG) regime applies when
$\gamma\gg 1$~\cite{de_rosi_thermodynamic_2017}. In the TG regime bosons start to become impenetrable and the Bose gas acquires fermionic properties.
Describing this correctly requires taking into account the discreteness of the system,
invalidating the gradient expansion used in this work.
Thus, we focus here on the MF regime, where both the ansatz and field representation are valid. 
Additional details can be found in  Appendix~\ref{app:TGregime}.

Fig.~\ref{fig:ZT_Thermo_1D}  shows the chemical potential, energy per particle and the microscopic sound velocity $c_s$ at zero temperature.
The first two are scaled in terms of the Fermi energy $E_F=\pi^2n_0^2/2m$. The sound velocity is shown in terms of the Fermi velocity $v_F=\pi n_0/m$. In the limit $\gamma\to\infty$
the system can be completely mapped into a Fermi gas and $v_s=v_F$.
We compare our results with \cite{lieb_exact_1963,lieb_exact_1963-1},
\begin{align}
\mu_0(\gamma\ll 1)=&\frac{n_0^2\gamma}{m}\left(1-\frac{\gamma^{1/2}}{\pi}\right),\label{eq:ZT_1D_mu_MF}\\
E/N(\gamma\ll 1)=&\frac{n_0^2\gamma}{2m}\left(1-\frac{4}{3\pi}\gamma^{1/2}\right),\label{eq:ZT_1D_EoN_MF}\\
v_s(\gamma\ll 1)=&\frac{n_0 \gamma^{1/2}}{m}\sqrt{1-\frac{\gamma^{1/2}}{2\pi}},\label{eq:ZT_1D_vs_MF}
\end{align}
where the first terms correspond to the Bogoliubov expressions, and the second to the first correction. Additionally, in order to compare the results with the expected behavior for larger values of $\gamma$, we show
analytical expressions for the TG regime \cite{de_rosi_thermodynamic_2017},
\begin{align}
\mu_0(\gamma\gg 1)=&\frac{\pi^2n_0^2}{2m}\left(1-\frac{16}{3\gamma}\right),\label{eq:ZT_1D_mu_TG}\\
E/N(\gamma\gg 1)=&\frac{\pi^2n_0^2}{6m}\left(1-\frac{4}{\gamma}\right),\label{eq:ZT_1D_EoN_TG}\\
v_s(\gamma\gg 1)=&\frac{\pi n_0}{m}\sqrt{1-\frac{8}{\gamma}},\label{eq:ZT_1D_vs_TG}
\end{align}
which contain first-order corrections to the expressions for the ideal Fermi gas.

\begin{figure}
	\centering
    \includegraphics[width=0.45\textwidth]{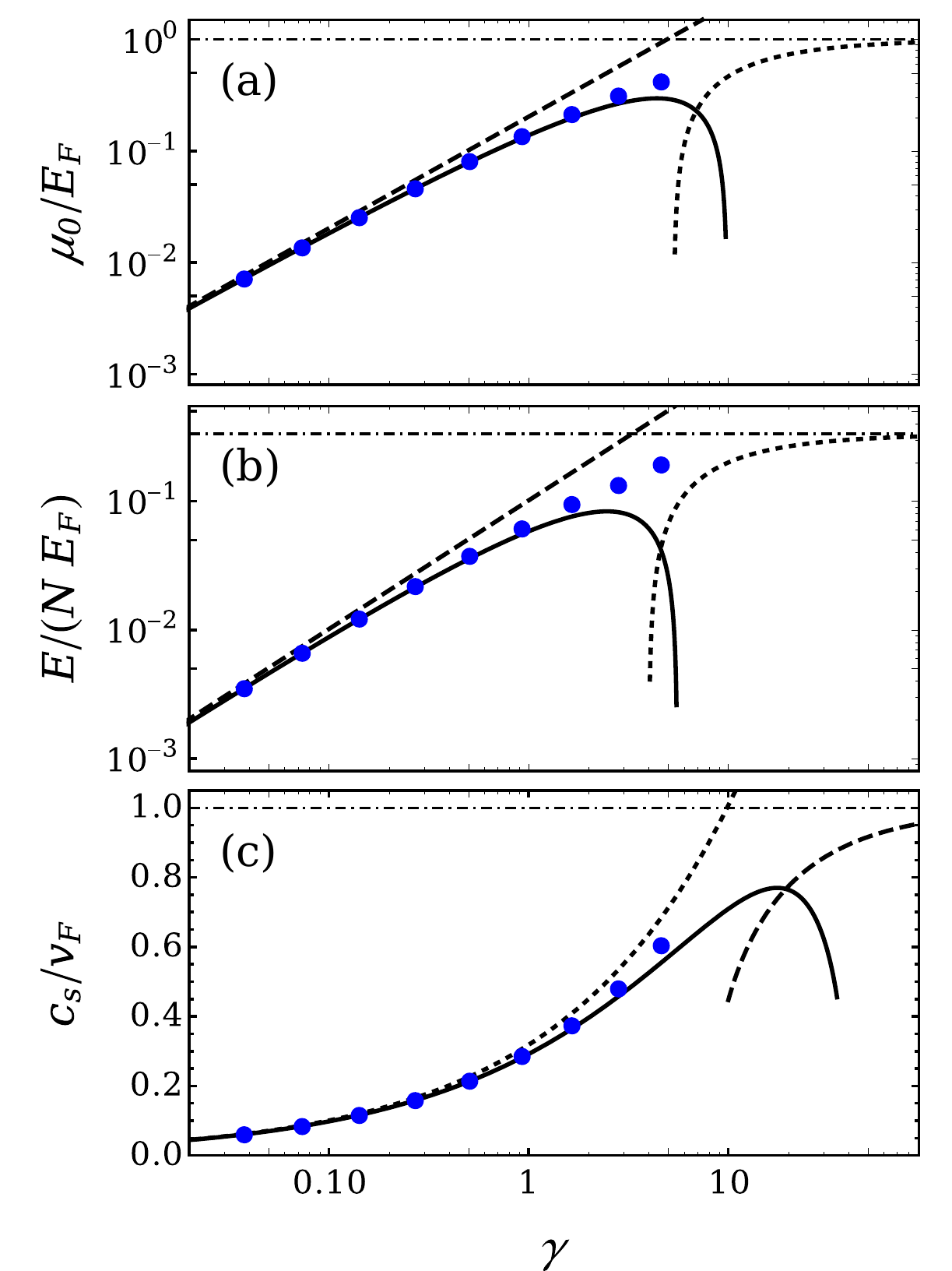}
    \caption{Chemical potential $\mu_0$ (a), energy per particle $E/N$ (b) and  microscopic sound velocity $c_s$ (c) 
    as functions of parameter $\gamma$ in one dimension  at $T=0$. 
    The blue circles are obtained using the interpolating representation.
    The dashed black lines are the Bogoliubov results for the MF regime, whereas the solid black lines include
    the first correction given in Eqs. (\ref{eq:ZT_1D_mu_MF}-\ref{eq:ZT_1D_vs_MF}).
    The dotted black lines correspond to expressions for the TG regime given in Eqs. (\ref{eq:ZT_1D_mu_TG}-\ref{eq:ZT_1D_vs_TG}),
    whereas the horizontal dash-dotted line to the free Fermi gas result.  } 
    \label{fig:ZT_Thermo_1D}   
\end{figure}

The figures show a remarkable agreement with the analytical results, following the corrections as $\gamma$ increases. 
This shows the usefulness of the interpolating representations as, to our knowledge, 
a successful description of the weakly-interacting one-dimensional Bose gas has not been obtained using other versions of the FRG.

At $\gamma\approx 1$ the system crosses into the TG regime, and thus the analytical expressions for the
MF regime are no longer valid. At $\gamma\approx 5$ the FRG flow becomes unstable, and we are unable to continue solving for higher values of $\gamma$.
Surprisingly, our results seem to be reasonable well into the crossover between these MF and TG regimes.

\section{Conclusions}  

In this work, we have studied the superfluid phases of weakly-interacting Bose gases in one, two and three dimensions using the functional renormalization group.
The boson fields are parametrized using the interpolating representation developed in our previous work \cite{isaule_application_2018}. This
makes it possible to solve the RG flow using a Cartesian representation at high momentum scales, and an amplitude-phase representation  at low scales. It allows us to
deal with Goldstone (phase) fluctuations in a natural manner, while also correctly integrating over the Gaussian UV fluctuations.
Our approach gives finite physical values for the wave-function renormalization and longitudinal mass, consistent with Popov's hydrodynamic effective action.

We also have developed a rather natural approach for calculating the thermodynamic properties of Bose gases. These are obtained
from the effective potential, as this is related to the grand canonical potential. The boson and entropy densities can be expressed as derivatives of the effective potential and their physical values can be found from the corresponding flow equations. Calculating the pressure 
can be difficult for FRG treatments, as it requires counter terms that are  difficult to fix numerically with frequency-independent regulators. 
By making the rather mild assumption that these counterterms are independent of temperature and chemical potential, we are able to calculate the pressure by integrating the physical values of the boson and entropy density, starting from the vacuum where the pressure vanishes.

To test this approach, we have fixed the initial conditions of the flow in terms of the $s$-wave scattering in each dimension
by renormalizing the interaction term $u_2$ in vacuum. This allowed us to study the properties of the system in terms
of the scattering length, the density and temperature, and  compare them with other approaches.

At zero temperature, our results for the thermodynamic properties 
are in agreement with the known corrections to the mean-field estimates in one and three dimensions, 
and with Monte-Carlo simulations in two dimensions. At finite temperature, we also 
find good agreement with results from simulations in three dimensions. This demonstrates the validity of our approach for calculating the pressure.
However, the superfluid phase in two dimensions at finite temperatures is not so well described. 
Here we find a continuous decrease of the superfluid fraction to zero instead of the sudden jump of the 
BKT transition. On the other hand, our values for the critical temperature in two dimensions are in excellent agreement with
the estimate given by Fisher and Hohenberg \cite{fisher_dilute_1988}, which also neglects vortex effects.
This indicates that vortices are an important missing part of our calculation in two dimensions, explaining why our results in the region around the BKT transition are unreliable.

Our results in one dimension are particularly interesting since, to our knowledge, 
it has not previously been possible to study this system with the FRG. 
Indeed, we are able to describe its weakly-interacting regime and even its approach to the strongly-interacting regime.
This further confirms the usefulness of the AP representation in studies of low-dimensional systems. 
The strongly-interacting regime is still not well described, but this is expected since the effects of
the periodic nature of this regime need to be included
(see Ref.~\cite{cazalilla_one_2011} for more details).

We conclude that the use of different field representations in the IR and UV regimes can  improve the description of Bose gases
and related systems using the FRG. However, further work is needed in order to obtain a numerically accurate description.
Also, the physics of inhomogeneous ground states in one and two dimensions is not described by our current truncation. 
Examples of these are vortex physics in two dimensions, and the periodic behavior of the strongly-interacting regime in one dimension, which could potentially be
included  using a modification of the AP representation as proposed by Cazalilla \cite{cazalilla_bosonizing_2004}.
The dependence of the interpolating representation on the truncation  should also be studied in future work, in particular the effects of
field-dependent wave-function and mass renormalization factors and   second-order time derivative terms.

Finally, one immediate extension of this work will be to apply our interpolating approach to Fermi gases in the BCS-BEC crossover regime. Since fermion pairs can be represented by boson fields,
Fermi gases show similar IR issues in their superfluid phases.  Furthermore, the AP representation should be particularly useful
for studying the pseudogap regime, as the quasicondensate density can be related to the magnitude of the pseudogap \cite{loktev_phase_2001}.

%\begin{acknowledgments}
\section*{Acknowledgments}
FI acknowledges funding from CONICYT Becas Chile under Contract No 72170497.
MCB and NRW are supported by the UK STFC under grants ST/L005794/1 and ST/P004423/1.
%\end{acknowledgments}

\appendix

\section{Ansatz and flow equations with the interpolating representation}
\label{app:Interpolating}

By inserting the definition (\ref{eq:InterpFields}) into  Eq.~(\ref{eq:Gammak}) we obtain a parametrization of $\Gamma$ for the broken phase in terms of the interpolating fields. It reads
\begin{multline}
\Gamma[\PHI]=-\int_x\Bigg[i \,b_k\big(A_k^2 \partial_\tau\ctheta-B_k\sin(\ctheta/\bk)\partial_\tau\sigma-A_kB_k\cos(\ctheta/\bk)\partial_\tau\ctheta\big)
+\frac{Z_\ctheta}{2m}A^2_k(\sigma)(\nabla\ctheta)^2\\+\frac{Z_\sigma(\ctheta)}{2m}(\nabla\sigma)^2+\frac{Y_m}{2m}\bk^2 
\times\Bigg(\frac{\sigma}{\bk}\bigg(\frac{\sigma}{\bk}+2\big(1-B_k\cos(\ctheta/\bk)\big)\bigg)(\nabla\sigma)^2\\+A^2_k(\sigma)B^2_k\sin^2(\ctheta/\bk)(\nabla\ctheta)^2
+2\Big(A_k(\sigma)-B_k\cos(\ctheta/\bk)\Big)\\ \times A_k(\sigma)B_k\cos(\ctheta/\bk)\nabla\sigma\nabla\ctheta \Bigg)+U(\rho,\mu)\Bigg],
\label{eq:Gammabk}
\end{multline}
where 
\begin{equation}
A_k(\sigma)=\left(1+\frac{\sigma}{\bk}\right), \, B_k=\left(1-\frac{\sqrt{\rho_0}}{\bk}\right).
\label{eq:AkBk}
\end{equation}
and
\begin{equation}
Z_\sigma(\ctheta)=Z_\ctheta+Y_m\bk^2\big(1-B_k\cos(\ctheta/\bk)\big)^2.
\end{equation}
The effective potential $U$ is defined in Eq.~(\ref{eq:Ueff}), where we take $u_1=0$ since we are working in the broken phase.
The density takes the form
\begin{equation}
\rho=\bk^2\left[A_k^2(\sigma)+B_k^2-2A_k(\sigma)B_k\cos(\ctheta/\bk)\right].
\label{eq:rhoInterp}
\end{equation}

The propagator evaluated at $\sigma=\ctheta=0$ is given by
\begin{equation}
\MG_b(q)=\frac{-1/2}{Z_\phi^2 q_0^2+E_{R,\sigma}(\Q) E_{R,\ctheta}(\Q)}\begin{pmatrix}
  E_{R,\ctheta}(\Q) & Z_\phi q_0 \\
  -Z_\phi q_0 & E_{R,\sigma}(\Q)
 \end{pmatrix},
 \label{eq:Gb}
\end{equation}
where
\begin{align}
E_{R,\sigma}(\Q)=&\frac{Z_\sigma}{2m}\Q^2-n_1 \delta\mu+2(u_2-n_2\delta\mu)\rho_0+R(\Q), \label{eq:ER_sigma}\\
E_{R,\ctheta}(\Q)=&\frac{Z_\ctheta}{2m}\Q^2-n_1 B_k \delta\mu +R(\Q),\label{eq:ER_ctheta}
\end{align}
are the regulated energies, with $\delta\mu=\mu-\mu_0$.

The flow equations for the $k$-dependent couplings and factors are extracted from field derivatives of Eq.~(\ref{eq:PawlowskEq}). At the level of truncation used in this work, Eqs.~(\ref{eq:Gammak},\ref{eq:Ueff}), they are: 
\begin{align}
2u_2\sqrt{\rho_0}\vrho_0=&\,\vGamma^{(1)}_{\sigma}\Big|_{\rho_0,\mu_0},\nonumber\\
-4\rho_0\vu2+2u_2\vrho_0=&\,\vGamma^{(2)}_{\sigma\sigma}\Big|_{\rho_0,\mu_0}, \nonumber\\
\vn0-n_1\vrho_0=&\,\partial_\mu\vGamma\Big|_{\rho_0,\mu_0}, \nonumber\\
2\sqrt{\rho_0}\vn1-2n_2\sqrt{\rho_0}\vrho_0=&\,\partial_\mu\left(\vGamma^{(1)}_{\sigma}\right)\Big|_{\rho_0,\mu_0}, \nonumber\\
4\rho_0\vn2+2\vn1-2n_2\vrho_0=&\,\partial_\mu\left(\vGamma^{(2)}_{\sigma\sigma}\right)\Big|_{\rho_0,\mu_0}, \nonumber\\
2\vZphi=&\partial_{p_0}\left(\partial_k \Gamma^{(2)}_{\sigma\ctheta}\right)\bigg|_{\rho_0,\mu_0,p=0}, \nonumber\\
-\frac{\vZctheta}{m}=&\,\partial_{\vP^2}\left(\vGamma^{(2)}_{\ctheta\ctheta}\right)\Big|_{\rho_0,\mu_0,p=0}, \nonumber\\
-\frac{\rho_0\vYm}{m}-\frac{\vZctheta}{m}=&\,\partial_{\vP^2}\left(\vGamma^{(2)}_{\sigma\sigma}\right)\Big|_{\rho_0,\mu_0,p=0}, \label{eq:FlowEqs_b}
\end{align}
The terms on the left-hand  sides of Eq.~(\ref{eq:FlowEqs_b}) arise from  derivatives of both terms on the 
left-hand-side of Eq.~(\ref{eq:PawlowskEq}) evaluated at $\rho=\rho_0$ and $\mu=\mu_0$. Similarly, the terms on the right-hand sides originate from both terms on the right-hand-side of Eq.~(\ref{eq:PawlowskEq}). The $\sigma$ and $\ctheta$ subscripts denote  derivatives with respect to those fields. 
The field derivatives are evaluated using the convention $\phi(q)=\int_q e^{i(\Q\cdot\X-q_0\tau)}\phi(x)$, where
\begin{equation}
\int_q=T \sum_n \int\frac{\ddQ}{(2\pi)^d},
\end{equation}
and $q_0=2\pi n T$ are the bosonic Matsubara frequencies.  

The evolution equations (\ref{eq:FlowEqs_b}) share the same diagrammatic structure as in our previous work \cite{isaule_application_2018}, 
with the difference that here we work with $p=(p_0,\vP)$ instead of a purely space momentum $\vP$.
Thus, we refer to that work for a detailed discussion and the explicit expressions for the flow equations and driving terms.

\section{Critical temperature of the two-dimensional Bose gas}
\label{app:Tc2D}

\begin{figure}
	\centering
    \includegraphics[width=0.45\textwidth]{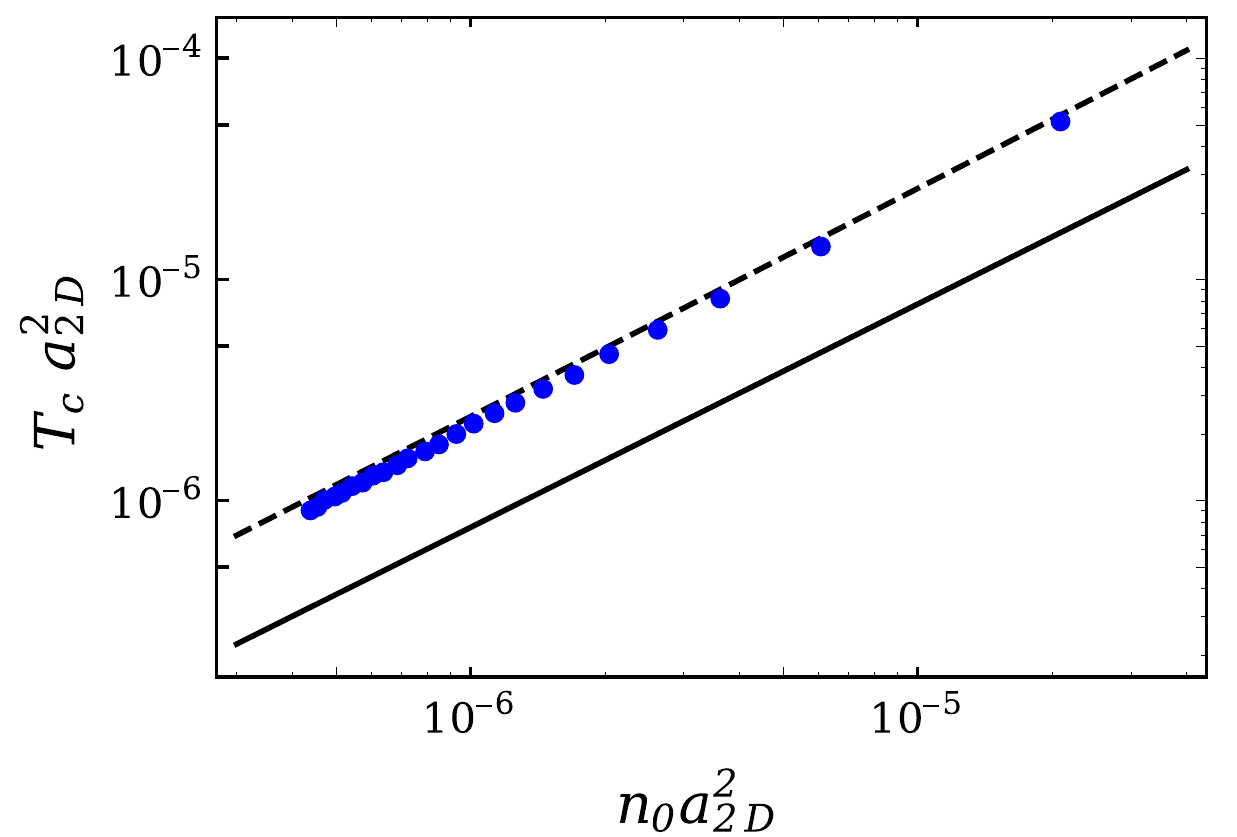}
    \caption{Critical temperature of the two-dimensional Bose gas as a function of $n_{0} a_{2D}^2$. The blue circles are obtained
    using the interpolating representation. The black dashed line is the estimate Eq.~(\ref{eq:Tc_2D_Fisher}), 
    and the solid black line  the BKT critical temperature (\ref{eq:Tc_BKT}).} 
    \label{fig:Tc_2D}   
\end{figure}

Fig.~\ref{fig:Tc_2D} shows the critical temperature for the superfluid phase transition as a function of the dimensionless
parameter $n_0 a_{2D}^2$. Following the estimate by Fisher and Hohenberg \cite{fisher_dilute_1988}, the critical temperature depends logarithmically on this parameter, 
\begin{equation}
T_c=\frac{2\pi n_0}{m}\frac{1}{\log\left(\log\left(1/n_0 a_{2D}^2\right)\right)}.
\label{eq:Tc_2D_Fisher}
\end{equation} 
This estimate, however, does not include vortex effects.
Vortices modify this expression to,
\begin{equation}
T_{BKT}=\frac{2\pi n_0}{m}\frac{1}{\log\left(\xi/4\pi\right)+\log\left(\log\left(1/n_0 a_{2D}^2\right)\right)},
\label{eq:Tc_BKT}
\end{equation}
where the constant $\xi=380$ is obtained from MC simulations \cite{prokofev_critical_2001}. As shown in our previous work, our approach predicts a noticeably higher critical temperature than that of the BKT transition. This is a result of our omission of vortex effects. Indeed, we observe that our results are
much closer to the estimate in Eq.~(\ref{eq:Tc_2D_Fisher}), as expected. 

\section{Tonks-Girardeau regime of the one-dimensional Bose gas}
\label{app:TGregime}

The one-dimensional Bose gas is weakly-interacting when $\gamma=-2/n_0 a_{1D}\ll 1$, and is strongly interacting when $\gamma \gg 1$, which
corresponds to the Tonks-Girardeau (TG) regime.
As discussed in Sec.~\ref{sec:Results}, when using $\alpha=1$ the flow starts becoming unstable around $\gamma \approx 5$. This is not unexpected,
as our approach is designed for the weakly-interacting regime.  However, we note that this instability point changes with $\alpha$, and
is not present for $\alpha\gtrsim 2$.

\begin{figure}
	\centering
    \includegraphics[width=0.45\textwidth]{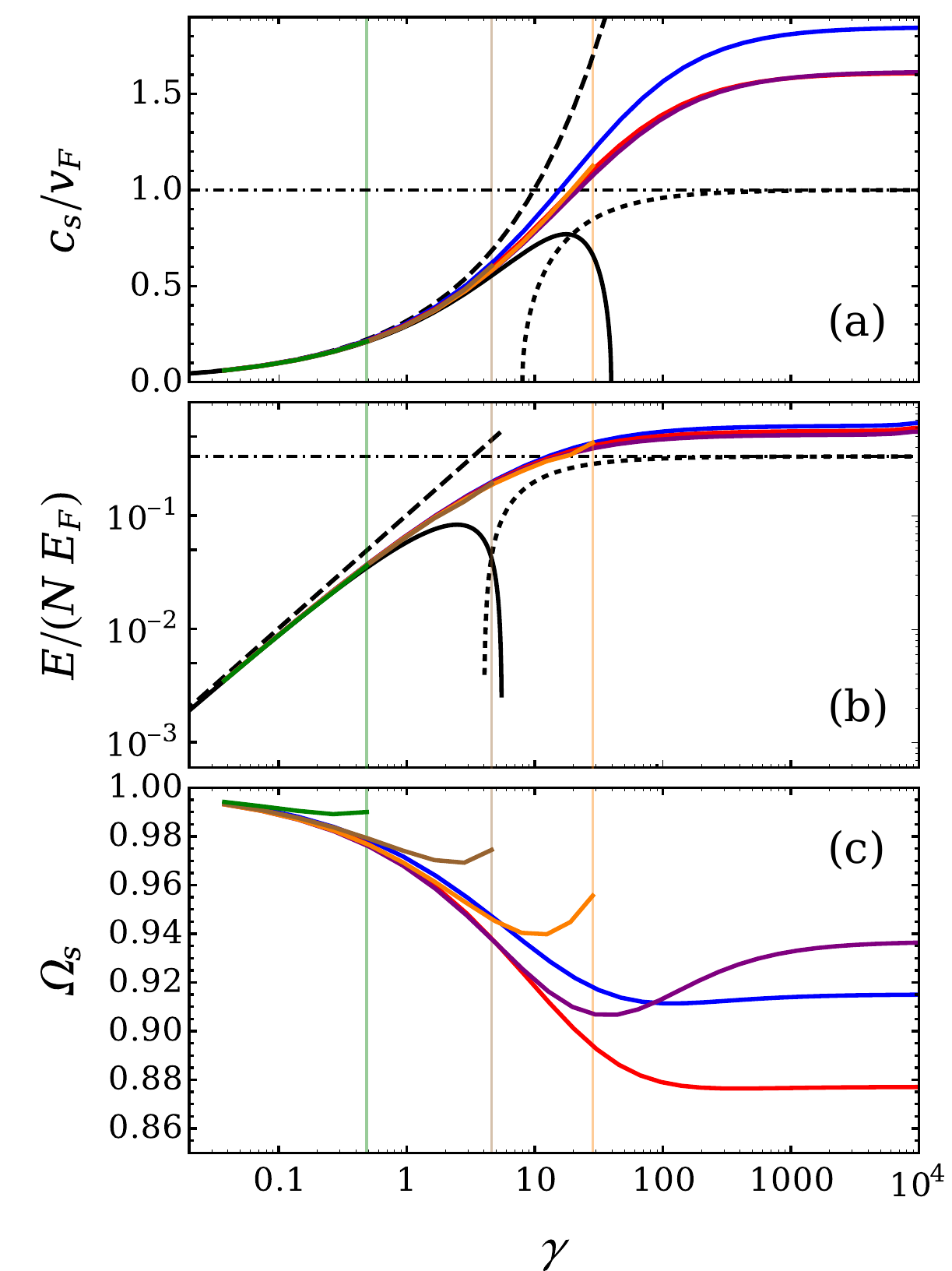}
    \caption{Ratio between microscopic sound velocity and Fermi velocity $c_s/v_F$ (a), energy per particle $E/N$ (b) and stiffness fraction
    $\Omega_s=\rho_s/n_0$ (c) as functions of parameter $\gamma$ in one dimension  at $T=0$.
 	The curves are obtained with $\alpha=4$ (blue), $\alpha=3$ (red), $\alpha=2$ (purple), $\alpha=1.5$ (orange), $\alpha=1$ (brown) and
 	$\alpha=0.5$ (green).
    The vertical lines show the value $\gamma$ where the flow starts becoming unstable for the corresponding value of $\alpha$.
    The dashed black lines are the Bogoliubov results for the MF regime, whereas the solid black lines include
    the first correction given in Eqs. ~(\ref{eq:ZT_1D_mu_MF}-\ref{eq:ZT_1D_vs_MF}).
    The dotted black lines correspond to expressions for the TG regime given in Eqs.~(\ref{eq:ZT_1D_mu_TG}-\ref{eq:ZT_1D_vs_TG}),
    whereas the horizontal dash-dotted line gives the free Fermi gas result.} 
    \label{fig:alphas_1D}   
\end{figure}

Fig.~\ref{fig:alphas_1D}  displays the sound velocity, energy per particle and stiffness fraction $\rho_s/n_0$ as a function of $\gamma$.
First, we note that for $\gamma\ll 1$ the different values of $\alpha$ result in indistinguishable results, thus our conclusion
from our previous work \cite{isaule_application_2018} that our approach is valid for values of $\alpha$ between $0.5$ and $2.0$ is still valid.
However, surprisingly  the results for $c_s$ and $E/N$ still seem to converge as the TG regime is approached, and for values of $\alpha$ between $2.0$ and $4.0$
they converge to similar results in the limit $\gamma\to\infty$, while deviations start showing for $\alpha\geq5$.
Still, as expected both the values for $c_s$ and $E/N$ are not correct. 
In the TG limit, where the system can be mapped onto a free Fermi gas, the sound velocity
should be equal to the Fermi velocity $v_F$, and the energy per particle to the corresponding energy of the Fermi gas $E_{TG}=\pi^2 n_0^2/6m$:
Both quantities are overestimated by around  $50\%$.

The deviations in the TG regime are more evident in the stiffness fraction, which shows a noticeable depletion as $\gamma$ increases.
As mentioned in Sec.~\ref{sec:Action} (see details in Ref.~\cite{cazalilla_one_2011}), this fraction should remain equal to one
for all $\gamma$. A similar depletion starts to appear in two and three dimensions when we try to solve the flow nearby the strongly interacting
regime, however the initial conditions constrain the flows to the regime where this depletion is small.

Despite the deviations, it is surprising that our approach gives some qualitative results in the TG regime. 
Indeed, our interpolating scheme can be seen as a first step to describe the one-dimensional Bose gas with the FRG.
As proposed by Cazalilla \cite{cazalilla_bosonizing_2004}, in the TG regime the AP representation should be modified to include the
periodic modulation of the two-body correlations
that emerge for a strongly repulsive Bose gas.

\bibliography{Biblio}

\end{document}